\documentclass[12pt]{amsart}
\usepackage[dvips,colorlinks=true,linkcolor=blue,urlcolor=red]{hyperref}
\usepackage{subfigure,caption2}
\usepackage{amssymb}
\usepackage{graphicx}
\usepackage[rflt]{floatflt}
\usepackage[francais]{babel} % si lettre en fran{{\c c}}ais
\numberwithin{equation}{section}
\newcommand{\smallpagebreak}{{\par\vspace{2 mm}\noindent}}

\newcommand{\dsize}{\textstyle}
\newcommand{\D}{\displaystyle}

\newcommand{\R}{{\mathbb R}}

\newcommand{\Z}{{\mathbb Z}}
\newcommand{\N}{{\mathbb N}}
\newcommand{\C}{{\mathbb C}}

\newcommand{\re}{{\rm Re}\,}
\newcommand{\im}{{\rm Im}\,}

\newcommand{\dist}{{\rm dist}\,}
\newcommand{\mes}{{\rm mes}\,}
\theoremstyle{plain}
\newtheorem{Th}{Th{\'e}or{\`e}me}[section]

\newtheorem{Pro}{Proposition}[section]
\newtheorem{Cor}{Corollaire}[section]
\theoremstyle{definition}
\newtheorem{Rem}{Remarque}[section]

\title{Op{\'e}rateurs de Schr{\"o}dinger quasi-p{\'e}riodiques adiabatiques :\\
  Interactions entre les bandes spectrales d'un op{\'e}rateur p{\'e}riodique}
\author{Alexandre Fedotov} \author{Fr{\'e}d{\'e}ric Klopp}
\address[Alexandre Fedotov]{D{\'e}partement de Physique Math{\'e}matique,
  Universit{\'e} d'{\'E}tat de Saint-P{\'e}tersbourg, 1, Ulia\-novskaja, 198904
  Saint-P{\'e}tersbourg -- Petrodvorets, Russie}
\email{\href{mailto:fedotov@mph.phys.spbu.ru}{fedotov@mph.phys.spbu.ru}}
\address[Fr{\'e}d{\'e}ric Klopp]{D{\'e}partement de Math{\'e}matique, Institut
  Galil{\'e}e, U.R.A 7539 C.N.R.S, Universit{\'e} de Paris-Nord, Avenue J.-B.
  Cl{\'e}ment, F-93430 Villetaneuse, France}
\email{\href{mailto: klopp@math.univ-paris13.fr}{klopp@math.univ-paris13.fr}}
\begin{document}
%% \french
%
\begin{abstract}
  This paper is devoted to the description of our recent results on
  the spectral behavior of one-dimensional adiabatic quasi-periodic
  Schr{\"o}dinger operators. The specific operator we study is a slow
  periodic perturbation of an incommensurate periodic Schr{\"o}dinger
  operator, and we are interested in energies where the perturbation
  creates a strong interaction between two consecutive bands of the
  background periodic operator. We describe the location of the
  spectrum and its nature and discuss the various new resonance
  phenomena due to the interaction of the spectral bands of the
  unperturbed periodic operator.
  \vskip.5cm
  \par\noindent   \textsc{R{\'e}sum{\'e}.}
  Dans cet article, nous d{\'e}crivons nos r{\'e}sultats r{\'e}cents sur la
  th{\'e}orie spectrale d'une classe d'op{\'e}rateurs de Schr{\"o}dinger
  quasi-p{\'e}riodiques adiabatiques sur la droite r{\'e}elle. Ces op{\'e}rateurs
  sont des perturbations p{\'e}riodiques lentes d'op{\'e}rateurs p{\'e}riodiques.
  Nous {\'e}tudions le spectre {\`a} des {\'e}nergies auxquelles la perturbation
  lente cr{\'e}e une interaction forte entre deux bandes spectrales
  cons{\'e}cutives de l'op{\'e}rateur p{\'e}riodique non perturb{\'e}.  Nous d{\'e}crivons
  le lieu et la nature du spectre ; nous nous int{\'e}ressons plus
  particuli{\`e}rement {\`a} diff{\'e}rents ph{\'e}nom{\`e}nes de r{\'e}sonance engendr{\'e}s par
  l'interaction entre les bandes spectrales de l'op{\'e}rateur p{\'e}riodique
  non perturb{\'e}.
\end{abstract}
\setcounter{section}{-1}
\maketitle
\section{Introduction}
\label{sec:intro}
Nous analysons le spectre de la famille d'op{\'e}rateurs de Schr{\"o}dinger
quasi-p{\'e}riodiques
\begin{equation}
  \label{family}
  H_{\zeta,\varepsilon}=-\frac{d^2}{dx^2}+V(x)+\alpha\cos(\varepsilon
  x+\zeta),
\end{equation}
agissant sur $L^2(\R)$. Nous supposerons que
\begin{description}
\item[(H1) ] $V$ est une fonction r{\'e}elle de la variable r{\'e}elle qui est
  $1$-p{\'e}riodique, de carr{\'e} localement int{\'e}grable et qui n'est pas
  constante ;
\item[(H2) ] $\varepsilon$ est un nombre positif qui, comme son nom le
  sugg{\`e}re, sera choisi petit ; de plus, nous supposerons que
  $2\pi/\varepsilon$ est irrationnel ;
\item[(H3) ] $\zeta$ est un param{\`e}tre r{\'e}el servant en particulier {\`a}
  indexer la famille d'{\'e}quations ;
\item[(H4) ] $\alpha$ est un param{\`e}tre strictement positif qui restera
  fix{\'e} dans la plus grande partie de l'expos{\'e}.
\end{description}
\noindent L'op{\'e}rateur~\eqref{family} est une perturbation lente de
l'op{\'e}rateur de Schr{\"o}dinger p{\'e}riodique
\begin{equation}
  \label{Ho}
  H_0=-\frac{d^2}{dx^2}+V\,(x)
\end{equation}
agissant sur $L^2(\R)$. Les r{\'e}sultats que nous allons d{\'e}crire sont
principalement tir{\'e}s de~\cite{Fe-Kl:04a,Fe-Kl:04b} et font suite {\`a} une
s{\'e}rie d'articles consacr{\'e}s {\`a} la m{\^e}me famille
d'op{\'e}rateurs~\cite{MR2003f:82043,Fe-Kl:01b,Fe-Kl:03a}. Ces travaux se
fondent sur l'analyse de l'{\'e}quation de monodromie pour la famille
d'op{\'e}rateurs quasi-p{\'e}riodiques~\eqref{family}. La matrice monodromie a
{\'e}t{\'e} d{\'e}finie dans~\cite{MR2003f:82043} ; c'est une g{\'e}n{\'e}ralisation non
triviale de celle utilis{\'e}e pour l'{\'e}tude des {\'e}quations aux diff{\'e}rences
finies, voir~\cite{Bu-Fe:96}. Pour {\'e}tudier cette matrice de
monodromie, nous d{\'e}velopp{\'e} une nouvelle m{\'e}thode asymptotique
dans~\cite{Fe-Kl:98c} et~\cite{Fe-Kl:03e}.
\smallpagebreak Dans nos travaux pr{\'e}c{\'e}dents, il a d{\'e}j{\`a} {\'e}t{\'e} observ{\'e} que
la position relative de la {\it fen{\^e}tre spectrale}
$\mathcal{F}(E):=[E-\alpha,E+\alpha]$ par rapport au spectre de $H_0$
joue un r{\^o}le crucial dans la d{\'e}termination des caract{\'e}ristiques
spectrales de $H_{\zeta,\varepsilon}$ {\`a} l'{\'e}nergie $E$.
\par Dans cet article, nous supposerons que cette position est celle
d{\'e}crite dans la figure~\ref{interactingfigure} ; dans cette figure, la
fen{\^e}tre spectrale est repr{\'e}sent{\'e}e au-dessus des bandes spectrales de
$H_0$. On se donne donc deux bandes spectrales cons{\'e}cutives s{\'e}par{\'e}es
par une unique lacune et on consid{\`e}re l'intervalle des {\'e}nergies telles
que $\mathcal{F}(E)$ couvre partiellement chacune des bandes
spectrales et couvre totalement la lacune entre ces deux bandes.
%
%%%%%%%%%%%%%%%%%%%%%%%%%%%%%%%%%%%%%%%%%%%%%%%%%%%%%%%%%%%%%%%%%%%%%% %
\begin{floatingfigure}[r]{3.5cm}
  \centering
  \includegraphics[bbllx=71,bblly=662,bburx=176,bbury=721,width=3.5cm]{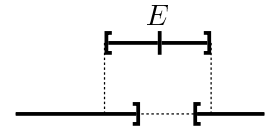}
  \caption{\og Bandes en interaction \fg}\label{interactingfigure}
\end{floatingfigure}
%%%%%%%%%%%%%%%%%%%%%%%%%%%%%%%%%%%%%%%%%%%%%%%%%%%%%%%%%%%%%%%%%%%%% %
%
D{\'e}crivons maintenant bri{\`e}vement que l'heuristique qui sous-tend nos
r{\'e}sultats ; le reste du texte sera d{\'e}volu {\`a} un expos{\'e} pr{\'e}cis de ces
r{\'e}sultats.
\par Soit $\mathbf{E}(\kappa)$ la relation de dispersion
associ{\'e}e {\`a} $H_0$ (voir la section~\ref{sec:le-quasi-moment}) ;
consid{\'e}rons les {\it vari{\'e}t{\'e}s caract{\'e}ristiques r{\'e}elle} et {\it
  complexe}, respectivement $\Gamma_\R$ et $\Gamma$, d{\'e}finies par
\begin{gather}
  \label{isoenr}
  \hskip-2cm\Gamma_\R:=\{(\kappa,\zeta)\in\R^2,\
  \mathbf{E}(\kappa)+\alpha\cdot \cos(\zeta)=E\},\\
  \label{isoen}
  \hskip-2cm\Gamma:=\{(\kappa,\zeta)\in\C^2,\
  \mathbf{E}(\kappa)+\alpha\cdot \cos(\zeta)=E\}.
\end{gather}
Ces courbes sont $2\pi$-p{\'e}riodiques dans les directions des $\kappa$
et des $\zeta$ ; elles sont d{\'e}crites dans la
partie~\ref{sec:iso-energy-curve-3}. Les composantes connexes de
$\Gamma_\R$ sont appel{\'e}es {\it branches r{\'e}elles} de la vari{\'e}t{\'e}
caract{\'e}ristique~\eqref{isoen}.
%
%%%%%%%%%%%%%%%%%%%%%%%%%%%%%%%%%%%%%%%%%%%%%%%%%%%%%%%%%%%%%%%%%%%%%%%%
\begin{floatingfigure}[r]{7.5cm}
  \centering
  \includegraphics[bbllx=71,bblly=566,bburx=247,bbury=721,width=7cm]{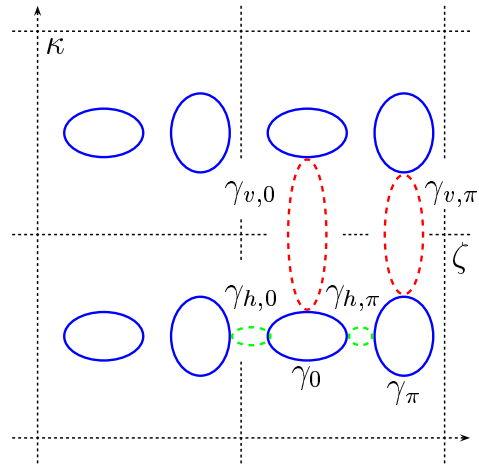}
  \caption{L'espace des phases adiabatique}\label{TIBMfig:actions}
\end{floatingfigure}
%%%%%%%%%%%%%%%%%%%%%%%%%%%%%%%%%%%%%%%%%%%%%%%%%%%%%%%%%%%%%%%%%%%%%%%%%
%
\vskip.1cm Consid{\'e}rons maintenant $J$, un intervalle d'{\'e}nergies tel
que, pour $E\in J$, notre hypoth{\`e}se sur la position relative de
$\mathcal{F}(E)$ par rapport au spectre de $H_0$ est v{\'e}rifi{\'e}e i.e. on
est dans le cas de la figure~\ref{interactingfigure}. Dans ce cas, la
courbe $\Gamma_\R$ consiste en une r{\'e}union infinie de composantes
connexes chacune hom{\'e}o\-morphe {\`a} un tore ; il y a exactement deux
telles composantes connexes par cellule de p{\'e}\-riodicit{\'e} (voir la
figure Fig.~\ref{TIBMfig:actions}). Sur cette figure, chaque carr{\'e}
correspond {\`a} une cellule de p{\'e}riodicit{\'e} du symbole
$(\zeta,\kappa)\mapsto\mathbf {E}(\kappa)+\alpha\cdot \cos(\zeta)$. Les
composantes con\-nexes de $\Gamma_\R$ sont repr{\'e}sent{\'e}es par une ligne
en trait plein ; on les notes $\gamma_0$ et $\gamma_\pi$ (une
d{\'e}finition pr{\'e}cise est donn{\'e}e
dans la section~\ref{sec:real-iso-energy-1}).\\
Les lignes en pointill{\'e}s repr{\'e}sentent des lacets dans $\Gamma$ qui
joignent certaines composantes con\-nexes de $\Gamma_\R$ (une
d{\'e}finition pr{\'e}cise est donn{\'e}e dans la
section~\ref{sec:complex-loops-1}) ; on distingue les lacets dits \og
horizontaux \fg~et ceux dits \og verticaux \fg.  On distingue deux lacets
horizontaux particuliers not{\'e}s $\gamma_{h,0}$ et $\gamma_{h,\pi}$ (ou,
plus pr{\'e}cis{\'e}ment, deux classes d'homotopie repr{\'e}sent{\'e}es chacune par un
tel lacet horizontal) ; le lacet $\gamma_{h,0}$ (resp.
$\gamma_{h,\pi}$) joint $\gamma_0$ {\`a} $\gamma_\pi-(2\pi,0)$ (resp.
$\gamma_0$ {\`a} $\gamma_\pi$). De m{\^e}me, on distingue deux lacets
verticaux particuliers not{\'e}s $\gamma_{v,0}$ et $\gamma_{v,\pi}$ ; le
lacet $\gamma_{v,0}$ (resp.  $\gamma_{v,\pi}$) joint $\gamma_0$ {\`a}
$\gamma_0+(0,2\pi)$ (resp. $\gamma_\pi$ {\`a} $\gamma_\pi+(0,2\pi)$) (les
d{\'e}finitions pr{\'e}cises sont donn{\'e}es dans la
section~\ref{sec:complex-loops-1}).
\par L'heuristique semi-classique standard sugg{\`e}re le comportement
spectral suivant. {\`A} chacun des lacets $\gamma_0$ et $\gamma_\pi$ est
associ{\'e}e une phase obtenue en int{\'e}grant la $1$-forme fondamentale de
la vari{\'e}t{\'e} $\Gamma$ le long de ce lacet ; on note ces phases
respectivement $\Phi_0=\Phi_0(E)$ and $\Phi_{\pi}=\Phi_{\pi}(E)$.
Chacune de ces phases d{\'e}finit une condition de quantification
\begin{equation}
  \label{eq:3}
  \frac{1}{\varepsilon}\Phi_0(E)=\frac\pi2+n\pi\text{ et }
  \frac{1}{\varepsilon}\Phi_\pi(E)=\frac\pi2+n\pi,\quad n\in\N.
\end{equation}
Chacune de ces conditions d{\'e}finit une suite de valeurs d'{\'e}nergie,
disons $(E_0^{(l)})_l$ et $(E_\pi^{(l')})_{l'}$, dans $J$. Pour
$\varepsilon$ suffisamment petit, le spectre de
$H_{\zeta,\varepsilon}$ dans $J$ devrait alors {\^e}tre situ{\'e}e dans un
voisinage de ces {\'e}nergies.
\par D'autre part, {\`a} chacun des lacets \og complexes \fg\ $\gamma_{0,h}$,
$\gamma_{\pi,h}$, $\gamma_{0,v}$ et $\gamma_{\pi,v}$ est associ{\'e}e une
action obtenue comme pr{\'e}c{\'e}demment en int{\'e}grant la $1$-forme
fondamentale de $\Gamma$ le long du dit lacet. Pour $a\in\{0,\pi\}$ et
$b\in\{v,h\}$, on appelle $S_{a,b}$ l'action associ{\'e}e {\`a}
$\gamma_{a,b}$. Lorsque $E\in\R$, toutes ces actions sont r{\'e}elles. On
oriente les contours d'int{\'e}gration de fa{\c c}on {\`a} ce qu'elles soient
toutes positives. Finalement, on d{\'e}finit des coefficients de tunnel
par
\begin{equation}
  \label{eq:4}
 t_{a,b}=e^{-S_{a,b}/\varepsilon},\quad a\in\{0,\pi\},\ b\in\{v,h\}.
\end{equation}
\smallpagebreak Lorsque, dans une cellule de p{\'e}riodicit{\'e}, la vari{\'e}t{\'e}
caract{\'e}ristique r{\'e}elle consiste en un unique tore
(voir~\cite{MR2003f:82043}), on sait que le spectre est contenu dans
une suite d'intervalles. Chacun de ces intervalles est voisin d'une
solution de la condition de quantification associ{\'e} au tore en
question. La longueur de ces intervalles est de l'ordre du plus grand
des coefficients de tunnel associ{\'e} {\`a} ce tore. La nature du spectre est
d{\'e}termin{\'e}e par le quotient du coefficient de tunnel vertical (i.e.
celui d'indice $v$) et du coefficient de tunnel horizontal (i.e. celui
d'indice $h$) selon la r{\`e}gle suivante : si le quotient est grand, le
spectre est singulier ; si le quotient est petit, le spectre est
absolument continu, voir~\cite{MR2003f:82043}.
\smallpagebreak Dans notre cas, il faut de plus tenir compte de
l'interaction entre les deux tores pr{\'e}sents dans la m{\^e}me cellule de
p{\'e}riodicit{\'e}. Comme dans le cas du \og double puits \fg\
(voir~\cite{MR85d:35085,He-Sj:84}), cet effet ne joue un r{\^o}le
important que lorsque deux {\'e}nergies, engendr{\'e}es par chacun des deux
tores, sont suffisamment proches l'une de l'autre. En fait, l'effet de
cette interaction est manifeste d{\`e}s que les {\'e}nergies quantifi{\'e}es
par~\eqref{eq:3} sont exponentiellement proches l'une de l'autre.
\smallpagebreak Consid{\'e}rons d'abord le cas des {\'e}nergies non
r{\'e}sonantes. Soit $E_0$ une {\'e}nergie satisfaisant {\`a} la condition de
quantification d{\'e}finie par $\Phi_0$.  Supposons de plus qu'une
distance d'ordre au moins $\varepsilon^n$ s{\'e}pare $E_0$ des valeurs
d'{\'e}nergie satisfaisant {\`a} la condition de quantification d{\'e}finie par
$\Phi_\pi$.\\
{\`A} ces {\'e}nergies, les {\'e}tats du syst{\`e}me ne \og voient \fg\ pas l'autre r{\'e}seau
de tores, ceux obtenus par translation de $\gamma_\pi$, pas plus
qu'ils ne \og sentent \fg\ les coefficients de tunnel associ{\'e}s {\`a} ces
tores, $t_{v,\pi}$. Tout se passe comme s'il n'existait qu'un seul
tore par cellule de p{\'e}riodicit{\'e}.  Pr{\`e}s de $E_0$, le spectre de
$H_{\zeta,\varepsilon}$ est situ{\'e} dans un intervalle de longueur de
l'ordre du maximum des coefficients de tunnel $t_{v,0}$ et
$t_h=t_{h,0}t_{h,\pi}$ (voir la section~\ref{sec:acti-integr-tunn-1}).
Ainsi, la nature du spectre est d{\'e}termin{\'e}e par le quotient
$t_{v,0}/t_h$.\\
Clairement, la situation est sym{\'e}trique pour des {\'e}nergies proches
seulement de solutions de la condition de quantification d{\'e}termin{\'e}e
par $\Phi_\pi$.\\
Dans le domaine des {\'e}nergies non r{\'e}sonantes, le spectre est contenu
dans deux suites d'intervalles exponentiellement petits ; pour chaque
suite, la nature du spectre est obtenue en comparant le coefficient de
tunnel vertical {\`a} celui horizontal pour le tore \og engendrant \fg\ cette
suite en suivant l'heuristique {\'e}tablie dans~\cite{MR2003f:82043}.
Comme les coefficients de tunnel pour les deux tores sont \og
ind{\'e}pendants \fg, il se peut que le spectre dans l'une des suites
d'intervalles soit singulier et que dans l'autre, il soit absolument
continu. Si tel est le cas, on obtient de nombreuses transitions
d'Anderson (i.e. de nombreux seuils s{\'e}parant du spectre absolument
continu de spectre singulier), voir la figure~\ref{fig:alt_spec}.
\vskip.2cm\noindent Dans le cas d'une {\'e}nergie $E$ r{\'e}sonante, i.e. qui
v{\'e}rifie \og presque \fg\ la condition de quantification pour chacun des
deux tores $\gamma_0$ et $\gamma_{\pi}$, on doit prendre en compte
l'interaction entre ces deux tores ; cet effet est analogue {\`a} celui
observ{\'e} dans le cas des puits multiples, en particulier, dans le cas
du double puits sym{\'e}trique. Dans le cas du double puits, on observe le
ph{\'e}nom{\`e}ne bien connu de r{\'e}pulsion de niveaux (\og splitting \fg), voir
\cite{MR85d:35085,MR81j:81010,He-Sj:84}.  Donc, dans notre cas, il
faut s'attendre {\`a} voir une r{\'e}pulsion des intervalles contenant le
spectre.  La nature du spectre est, elle aussi, affect{\'e}e par les
r{\'e}sonances. Pour simplifier la discussion, supposons que $t_{v,0}=
t_{v,\pi}\equiv t_v$. Dans le cas o{\`u} $E$ v{\'e}rifie en m{\^e}me temps les
deux conditions de quantification~\eqref{eq:3}, les deux r{\'e}seaux de
tores forment un seul r{\'e}seau. Pour ce nouveau r{\'e}seau, les tores des
deux types jouent le m{\^e}me r{\^o}le, et le coefficient de tunnel \og
horizontal \fg\ est {\'e}gal {\`a} $t_{h,0}=t_{h,\pi}=\sqrt{t_h}$.  Donc, la
nature du spectre est d{\'e}finie par le quotient $t_v/\sqrt{t_h}$ qui est
exponentiellement petit par rapport au quotient $t_v/t_h$ d{\'e}finissant
la nature du spectre dans le cas des {\'e}nergies non r{\'e}sonantes. Donc, si
dans le cas \og non r{\'e}sonant \fg, le spectre {\'e}tait singulier, dans le cas
\og r{\'e}sonant \fg, il peut devenir absolument continu. On peut donc trouver
des transitions d'Anderson dues aux r{\'e}sonances, voir la
figure~\ref{fig:res_tun}. En g{\'e}n{\'e}ral, on peut dire que les {\'e}tats
tendent {\`a} devenir moins localis{\'e}s. On peut donc s'attendre {\`a} voir des
variations tr{\`e}s marqu{\'e}es dans le comportement de l'exposant de
Liapounoff (qui mesure la vitesse de d{\'e}croissance des fonction
propres).
\smallpagebreak Dans le cas des {\'e}nergies r{\'e}sonantes, on voit donc un
ensemble de ph{\'e}nom{\`e}nes spectraux beaucoup plus riche que dans le cas
des {\'e}nergies non r{\'e}sonantes, ph{\'e}nom{\`e}nes dont la description fait
l'objet de cet article.
\smallpagebreak Il y a, en particulier, un jeu subtil entre la
r{\'e}pulsion et le comportement de l'exposant de Liapounoff. On voit une
r{\'e}pulsion forte sur la plupart des intervalles r{\'e}sonants sur lesquels
l'exposant de Liapounoff s'annule : dans ce cas-l{\`a}, il reste toujours
une lacune entre les intervalles r{\'e}sonants. Ces intervalles
contiennent majoritairement du spectre absolument continu et les {\'e}tats
correspondants sont \og {\'e}tendus \fg.\\
Pour les intervalles o{\`u} l'exposant de Liapounoff est positif et
d'ordre $1$, la r{\'e}pulsion est n{\'e}gligeable : la lacune entre les
intervalles r{\'e}sonants peut dispara{\^\i}tre. Remarquons que, sur les
intervalles o{\`u} le Liapounoff est positif, il n'y a que du spectre
singulier et les {\'e}tats correspondants sont \og localis{\'e}s \fg.\\
Finalement, on voit une r{\'e}pulsion des intervalles sur lesquels
l'exposant de Liapounoff devient anormalement petit (les {\'e}tats dans
ces intervalles ne sont que \og l{\'e}g{\`e}rement \fg\ localis{\'e}s) : dans ce cas
aussi, il reste toujours une petite lacune entre les intervalles
r{\'e}sonants.
\section{L'op{\'e}rateur p{\'e}riodique}
\label{sec:periodic-operator}
Dans cette partie, nous d{\'e}crivons la th{\'e}orie spectrale de l'op{\'e}rateur
de Schr{\"o}dinger p{\'e}riodique $H_0$ (agissant sur $L^2(\R)$) d{\'e}fini
dans~\eqref{Ho} (pour plus de pr{\'e}cisions ainsi que des preuves le
lecteur pourra se reporter {\`a}~\cite{Eas:73,MR2002f:81151}).
\subsection{Son spectre}
\label{sec:son-spectre}
Le spectre de l'op{\'e}rateur~\eqref{Ho} consiste en une r{\'e}union
d'intervalle de l'axe r{\'e}el $[E_{2n+1},\,E_{2n+2}]$, $n\in\N$, pour
lesquels
\begin{gather*}
  E_1<E_2\le E_3<E_4\dots E_{2n}\le E_{2n+1}<E_{2n+2}\le \dots\,,\\
  E_n\to+\infty,\quad n\to+\infty.
\end{gather*}
Ce spectre est purement absolument continu. Les points
$(E_{j})_{j\in\N}$ sont les valeurs propres de l'op{\'e}rateur obtenu en
consid{\'e}rant le polyn{\^o}me diff{\'e}rentiel~\eqref{Ho} agissant sur
$L^2([0,2])$ avec des conditions au bord p{\'e}riodiques. Les intervalles
introduits ci-dessus sont les {\it bandes spectrales}, et les
intervalles $(E_{2n},\,E_{2n+1})$, $n\in\N^*$, sont appel{\'e}s {\it
  lacunes spectrales}. Lorsque $E_{2n}<E_{2n+1}$, on dit que la
$n$-i{\`e}me lacune est {\it ouverte}, et, lorsque $[E_{2n-1},E_{2n}]$ est
s{\'e}par{\'e} du reste du spectre par des lacunes ouvertes, la $n$-i{\`e}me bande
est dite {\it isol{\'e}e}.
\smallpagebreak Dor{\'e}navant, pour simplifier notre expos{\'e}, nous
supposerons que
\begin{description}
\item[(O)] toutes les lacunes du spectre de $H_0$ sont ouvertes.
\end{description}
\subsection{La quasi-impulsion de Bloch}
\label{sec:le-quasi-moment}
Soit $x\mapsto\psi(x,E)$ une solution non triviale de l'{\'e}quation de
Schr{\"o}dinger p{\'e}riodique $H_0\psi=E\psi$ pour laquelle il existe
$\mu\in\C^*$ telle que $\psi\,(x+1,E)=\mu \,\psi\,(x,E)$, $\forall
x\in\R$.  On dit que $\psi$ est {\it solution de Bloch} de l'{\'e}quation,
et que $\mu =\mu (E)$ est {\it le multiplicateur de Floquet} associ{\'e} {\`a}
$\psi$. On peut {\'e}crire $\mu (E)=\exp(ik(E))$ ; la fonction $E\mapsto
k(E)$ est appel{\'e}e {\it quasi-impulsion de Bloch}. La solution de Bloch
$\psi$ s'{\'e}crit donc $\psi(x,E)=e^{\dsize ik(E)x}p(x,E)$ o{\`u} $x\mapsto
p(x,E)$ est une fonction $1$-p{\'e}riodique.
\smallpagebreak La fonction $E\mapsto k(E)$ est analytique et
multi-valu{\'e}e ; ses points de branchement sont les points $E_1$, $E_2$,
$E_3$, $\dots$, $E_n$, $\dots$. Ils sont tous de type \og racine carr{\'e}
\fg.
\smallpagebreak La relation de dispersion $k\mapsto{\bold E}(k)$ est
l'inverse de la quasi-impulsion de Bloch.
\smallpagebreak Soit $D$, un domaine simplement connexe ne contenant
pas de points de branchement du quasi-impulsion de Bloch $k$. Sur $D$,
fixons $k_0$, une d{\'e}termination continue (donc analytique) de $k$.
Toutes les autres d{\'e}terminations continues de $k$ sur $D$ sont
d{\'e}crites par
\begin{equation}\label{eq:55}
   k_{\pm ,l}(E)=\pm k_0(E)+2\pi l,\quad l\in\Z.
\end{equation}
Soit $\C_+$, le demi-plan complexe sup{\'e}rieur. Il existe $k_p$, une
d{\'e}termination analytique de $k$ qui envoie de fa{\c c}on conforme $\C_+$
sur le quadrant $\{k\in\C;\ \im k> 0,\,\,\re k> 0\}$ coup{\'e} le long
d'intervalles compacts du type $\pi l+i I_l$, $l=1,2,3\dots$,
$I_l\subset\R$. La d{\'e}termination $k_p$ se prolonge continuement {\`a}
$\C_+\cup\R$. Elle est r{\'e}elle et croissante le long du spectre de
$H_0$ ; elle envoie la bande spectrale $[E_{2n-1}, E_{2n}]$ sur
l'intervalle $[\pi(n-1),\pi n]$. Sur les lacunes ouvertes, $\re k_p$
est constante, et $\im k_p$ est positive et admet exactement un
maximum ; ce dernier est non d{\'e}g{\'e}n{\'e}r{\'e}.
\smallpagebreak Pour obtenir davantage de renseignements sur la
quasi-impulsion de Bloch, le lecteur pourra
consulter~\cite{Ke-Mo:75,MR55:761,MR2002f:81151}
\subsection{La fonctionnelle  $\Lambda_n(V)$}
\label{sec:le-coefficient-theta}
Nous d{\'e}crivons maintenant une fonctionnelle du potentiel p{\'e}riodique
$V$ importante pour l'{\'e}tude spectrale de~\eqref{family}.\\
Pour $E\in\C_+$, il  existe deux solutions de Bloch $x\mapsto\psi_\pm(x,E)$ de
la forme
\begin{equation*}
 \psi_\pm(x,E)=e^{\dsize\pm ik_p(E)x}p_\pm(x,E)
\end{equation*}
o{\`u} $x\mapsto p_\pm(x,E)$ sont des fonctions $1$-p{\'e}riodiques. On peut
normaliser les solutions de Bloch par la condition $\psi_\pm(0,E)=1$.
Les solutions normalis{\'e}es par cette condition sont d{\'e}finie d'une fa{\c c}on
unique ; elles sont analytique en $E\in \C_+$.
Consid{\'e}rons la fonction
\begin{equation}
  \label{eq:18}
  \omega(E)=-\frac{\int_0^1 p_-(x,E)\,
    \frac{\partial p_+}{\partial E}(x,E)dx}
   {\int_0^1 p_-(x,\, E)\,p_+(x,E)dx}.
\end{equation}
Elle a {\'e}t{\'e} {\'e}tudi{\'e}e dans~\cite{Fe-Kl:03e}. D{\'e}finie sur $\C_+$, elle
peut {\^e}tre prolong{\'e}e analytiquement {\`a} $\C$ priv{\'e} du compl{\'e}mentaire du
spectre de $H_0$ dans l'axe r{\'e}el, c'est-{\`a}-dire {\`a} $\C$ priv{\'e} des
intervalles $(-\infty, E_1]$ et $[E_{2j}, E_{2j+1}]$, $j\in\N^*$.
\smallpagebreak Soit $L_n=]E_{2n},E_{2n+1}[$, une lacune du spectre de
$H_0$.  Alors, {\`a} cette lacune, on associe le nombre $\Lambda_n$ d{\'e}fini
par
\begin{equation}
  \label{theta-Lambda}
  \Lambda_n=\frac12\left(\theta_n+\frac1{\theta_n}\right),\quad
  \theta_n=\exp\left(\oint_{g_n}\omega(E)dE\right).
\end{equation}
o{\`u} $g_n$ est le lacet entourant la lacune $L_n$.
\smallpagebreak Fixons $n$ et consid{\'e}rons $\Lambda_n$ comme une
fonctionnelle du potentiel $1$-p{\'e}riodique $V$. Pour cela, on identifie
les fonctions $1$-p{\'e}riodiques de carr{\'e} localement int{\'e}grable {\`a}
$L^2([0,1],\R)$ et on d{\'e}montre le r{\'e}sultat suivant
\begin{Th}[\cite{Fe-Kl:04b}]
  \label{w:prop}
  La fonctionnelle $V\mapsto\Lambda_n(V)$ a les propri{\'e}t{\'e}s suivantes:
  \begin{itemize}
  \item pour tout $V\in L^2([0,1],\R)$, on a $\Lambda_n(V)\ge 1$ ;
  \item sur un ouvert dense de $L^2([0,1],\R)$, on a $\Lambda_n(V)>1$
    ;
  \item si $V$ est pair, alors $\Lambda_n(V)=1$ ;
  \item elle est invariante par l'action du groupe des translations
    $(\tau_s)_{s\in\R}$ o{\`u} $\tau_s(V)(\cdot)=V(\cdot+s)$.
  \end{itemize}
\end{Th}
\noindent On peut donc consid{\'e}rer $\Lambda_n(V)$ comme une mesure de
la parit{\'e} du potentiel $V$.
\section{Une hypoth{\`e}se \og g{\'e}om{\'e}trique \fg\ sur la r{\'e}gion d'{\'e}nergie}
\label{sec:main-assumption-w}
Nous pouvons maintenant d{\'e}crire la r{\'e}gion des {\'e}nergies o{\`u} nous allons
{\'e}tudier le spectre de~\eqref{family}.
\smallpagebreak La {\it fen{\^e}tre spectrale centr{\'e}e en $E$} est
l'intervalle $\mathcal{F}(E)=[E-\alpha,E+\alpha]$. C'est l'image de
$\R$ par l'application $\zeta\mapsto E-\alpha\cos(\zeta)$.
\smallpagebreak Nous consid{\'e}rons les intervalles compacts $J\subset
\R$ tels que, pour $E\in J$, la fen{\^e}tre $\mathcal{F}(E)$ contient
exactement deux bords de bandes spectrales de $H_0$ provenant de deux
bandes distinctes (voir la figure~\ref{interactingfigure})
c'est-{\`a}-dire nous supposons qu'il existe $n\in\N^*$ tels que, pour
$E\in J$, on a
\begin{description}
  \label{TIBMcondition}
\item[(BEI) ]
  $[E_{2n},E_{2n+1}]\subset\dot{\mathcal F}(E)$ et  ${\mathcal
    F}(E)\subset]E_{2n-1},E_{2n+2}[$.
\end{description}
o{\`u} $\dot{\mathcal F}(E)$ d{\'e}signe l'int{\'e}rieur de $\mathcal{F}(E)$.
\begin{Rem}
  \label{rem:4}
  Comme toutes les lacunes spectrales de $H_0$ sont ouvertes, comme
  leur longueur tend vers $0$ et que la longueur des bandes spectrales
  tend vers l'infini, on voit que, pour toute valeur non nulle de
  $\alpha$, l'hypoth{\`e}se (BEI) est v{\'e}rifi{\'e}e dans toute lacune d'{\'e}nergie
  assez grande ; il suffit que cette lacune soit de longueur
  inf{\'e}rieure {\`a} $2\alpha$.
\end{Rem}
Dans toute la suite, $J$ est un intervalle compact v{\'e}rifiant (BEI) et
$\Lambda_n$ d{\'e}signe le coefficient d{\'e}fini dans la
section~\ref{sec:le-coefficient-theta} pour la lacune
$]E_{2n},E_{2n+1}[$ de l'hypoth{\`e}se (BEI).
\section{L'impulsion complexe}
\label{sec:iso-energy-curve}
Pour d{\'e}crire nos r{\'e}sultats, il nous faut introduire l'impulsion
complexe, quantit{\'e} centrale de la m{\'e}thode WKB complexe adiabatique
construite dans~\cite{Fe-Kl:98c,Fe-Kl:03e}. Pour cela, nous fixons une
valeur d'{\'e}nergie $E$ dans $J$.
\smallpagebreak On d{\'e}finit {\it l'impulsion complexe}
$\zeta\mapsto\kappa(\zeta)$ par
\begin{equation}
  \label{complex-mom}
  \kappa(\zeta)=k(E-\alpha\cos(\zeta)).
\end{equation}
Remarquons que~\eqref{complex-mom} est {\'e}quivalent {\`a} la
relation~\eqref{isoen}. {\`A} l'instar de $k$, $\kappa$ est analytique et
multi-valu{\'e}e. Les points de branchements de $k$ {\'e}tant les points
$(E_i)_{i\in\N}$, ceux de $\kappa$ v{\'e}rifient
\begin{equation}
  \label{BPCM}
  E-\alpha\cos(\zeta)=E_n,\ n\in\N.
\end{equation}
L'ensemble de ces points de branchement est sym{\'e}trique par rapport {\`a}
l'axe r{\'e}el et est $2\pi$-p{\'e}riodique en $\zeta$. Comme $E$ est r{\'e}el,
tous les points de branchements sont situ{\'e}s $\arccos(\R)$, la
pr{\'e}-image de $\R$ par $\alpha\cdot\cos$. L'ensemble $\arccos(\R)$ est
form{\'e} de l'axe r{\'e}el et de tous les translat{\'e}s de l'axe imaginaire par
un multiple entier de $\pi$.
\smallpagebreak {\`A} l'instar des points de branchement de la
quasi-impulsion de Bloch, les points de branchements de $\kappa$ sont
de type \og racine carr{\'e} \fg\ sauf s'ils co{\"\i}ncident avec l'un des points
$l\pi$, $l\in\Z$ ; sous l'hypoth{\`e}se (BEI), cette derni{\`e}re possibilit{\'e}
est exclue.
\smallpagebreak D{\'e}crivons maintenant ces points de branchement. Comme
le cosinus est analytique r{\'e}el, pair et $2\pi$-p{\'e}riodique, il suffit
de d{\'e}crire les points de branchement dans la bande $\{\zeta;\
\im\zeta\geq0,\ 0\leq\re\zeta\leq\pi\}$ ; tous les autres points de
branchement sont ensuite obtenus par les translations d'argument
$2n\pi$ (pour $n$ entier relatif), par r{\'e}flexion par rapport {\`a} l'axe
r{\'e}el et par sym{\'e}trie par rapport {\`a} l'origine.
%
%%%%%%%%%%%%%%%%%%%%%%%%%%%%%%%%%%%%%%%%%%%%%%%%%%%%%%%%%%%%%%%%%%%%%% %
\begin{floatingfigure}[r]{7cm}
  \centering
  \includegraphics[bbllx=71,bblly=606,bburx=275,bbury=721,width=7cm]{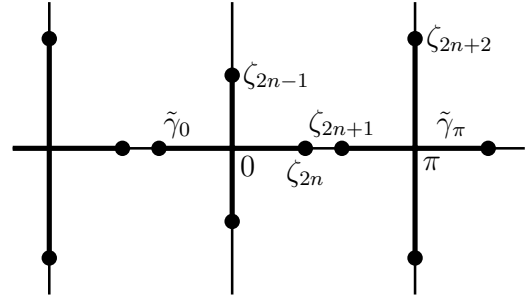}
  \caption{Les points de branchements}\label{dessin3a-tibm}
\end{floatingfigure}
%%%%%%%%%%%%%%%%%%%%%%%%%%%%%%%%%%%%%%%%%%%%%%%%%%%%%%%%%%%%%%%%%%%%% %
%
La quasi-impulsion complexe $\kappa$ a exactement deux points de
branchements dans $[0,\pi[$. En accord avec~\eqref{BPCM}, nous les
notons $\zeta_{2n}$ et $\zeta_{2n+1}$. Ils v{\'e}rifient
$\zeta_{2n}<\zeta_{2n+1}$ (voir la figure~\ref{dessin3a-tibm}). Comme
$\zeta\mapsto\cos(\zeta)$ est sym{\'e}trique par rapport {\`a} la droite
$\re\zeta=\pi$, les points de branchements de $\kappa$ dans
$[\pi,2\pi[$ sont les points $2\pi-\zeta_{2n}$ et
$2\pi-\zeta_{2n+1}$.\\
Les points de branchement hors de l'axe r{\'e}el sont situ{\'e}s sur les
droites $n\pi+i\R$, $n\in\{0,1\}$. En accord avec~\eqref{BPCM}, ceux
sur $i\R$ sont nomm{\'e}s $(\zeta_j)_{1\leq j\leq 2n-1}$, et on les
ordonne de fa{\c c}on {\`a} ce que $0<\im\zeta_{2n-1}<\cdots<\im\zeta_1$. De
m{\^e}me, ceux sur $\pi+i\R$ seront not{\'e}s $(\zeta_j)_{2n+2\leq j}$
ordonn{\'e}s de fa{\c c}on {\`a} ce que $0<\im\zeta_{2n+2}<\im\zeta_{2n+3}<\cdots$
(voir la figure~\ref{dessin3a-tibm}).
\section{La vari{\'e}t{\'e} caract{\'e}ristique}
\label{sec:iso-energy-curve-3}
Nous d{\'e}crivons maintenant $\Gamma_\R$ et $\Gamma$ d{\'e}finies
en~\eqref{isoenr} et~\eqref{isoen}. Nous ne d{\'e}crirons pas la totalit{\'e}
de la topologie de $\Gamma$ mais seulement certains lacets qui sont
utiles dans notre {\'e}tude.
\subsection{La vari{\'e}t{\'e} caract{\'e}ristique r{\'e}elle}
\label{sec:real-iso-energy-1}
Cette vari{\'e}t{\'e} est $2\pi$-p{\'e}riodique en $\kappa$ et en $\zeta$. Elle
est sym{\'e}trique par rapport aux droites $\zeta=\pi n$ et $\kappa=\pi
m$ pour $n$ et $m$ entiers relatifs.\\
D{\'e}crivons la partie de $\Gamma_\R$ contenue dans une cellule de
p{\'e}riodicit{\'e}. L'application ${\mathcal E}:\,\,\zeta\mapsto
E-\alpha\cos(\zeta)$ envoie l'intervalle $[\zeta_{2n+1},\pi]$ dans la
bande spectrale $[E_{2n+1},E_{2n+2}]$, l'intervalle $[0,\zeta_{2n}]$
dans la bande spectrale $[E_{2n-1},E_{2n}]$, et l'intervalle
$]\zeta_{2n},\zeta_{2n+1}[$ dans la lacune $]E_{2n},E_{2n+1}[$. Ainsi,
$\kappa_p(\zeta)=k_p({\mathcal E}(\zeta))$ est r{\'e}elle sur les
intervalles $[\zeta_{2n+1},\pi]$ et $[0,\zeta_{2n}]$, et
elle est de partie imaginaire positive sur $]\zeta_{2n},\zeta_{2n+1}[$.\\
Cela implique que, dans la bande $\{0\leq\re\zeta\leq\pi\}$, toutes
les composantes connexes de $\Gamma_\R$ sont situ{\'e}es \og au-dessus \fg\
des intervalles $[\zeta_{2n+1},\pi]$ et $[0,\zeta_{2n}]$.\\
Les graphes de $\kappa_p$ sur ces intervalles sont contenus dans
certaines composantes connexes de $\Gamma_\R$. Remarquons que
$\kappa_p$ est monotone sur chacun des intervalles
$[\zeta_{2n+1},\pi]$ et $[0,\zeta_{2n}]$ et que
\begin{equation*}
  \pi(n-1)<\kappa_p(0)<\kappa_p(\zeta_{2n})=\pi
  n=\kappa_p(\zeta_{2n+1})<\kappa_p(\pi)<\pi (n+1).
\end{equation*}
Les sym{\'e}tries de $\Gamma_\R$ impliquent que chacun de ces graphes est
un \og quart \fg\ d'une des deux composantes connexes de $\Gamma_\R$. La
composante connexe de $\Gamma_\R$ correspondant au graphe de
$\kappa_p$ sur $[\zeta_{2n+1},\pi]$, appelons la $\gamma_\pi$, est
sym{\'e}trique par rapport {\`a} la droite $\zeta=\pi$ ; celle correspondant
au graphe de $\kappa_p$ sur $[0,\zeta_{2n}]$, appelons la $\gamma_0$,
est sym{\'e}trique par rapport {\`a} la droite $\zeta=0$. Les deux composantes
sont sym{\'e}triques par rapport {\`a} la droite $\kappa=\pi n$ et elles sont
hom{\'e}omorphes au cercle.\\
Toutes les autres composantes connexes de la vari{\'e}t{\'e} caract{\'e}ristique
r{\'e}elle sont obtenues en translatant $\gamma_0$ et $\gamma_{\pi}$ d'un
multiple entier de $2\pi$ dans les directions verticales ou horizontales
(i.e. en leur appliquant les translations de vecteur $(2\pi n,2\pi m)$
pour $(n,m)\in\Z^2$). Sur la figure~\ref{TIBMfig:actions}, nous avons
repr{\'e}sent{\'e} en trait plein quelques p{\'e}riodes d'un exemple de vari{\'e}t{\'e}
caract{\'e}ristique r{\'e}elle.\\
Chaque cellule de p{\'e}riodicit{\'e} contient exactement deux composantes
connexes de $\Gamma_\R$.
\subsection{Les lacets sur la vari{\'e}t{\'e} caract{\'e}ristique complexe}
\label{sec:complex-loops-1}
Soit $\Pi:\,\Gamma\to\C$, la projection $\Pi(\kappa,\zeta)=\zeta$.
%
%%%%%%%%%%%%%%%%%%%%%%%%%%%%%%%%%%%%%%%%%%%%%%%%%%%%%%%%%%%%%%%%%%%%%% %
\begin{floatingfigure}[r]{7cm}
  \centering
  \includegraphics[bbllx=71,bblly=606,bburx=275,bbury=721,width=7cm]{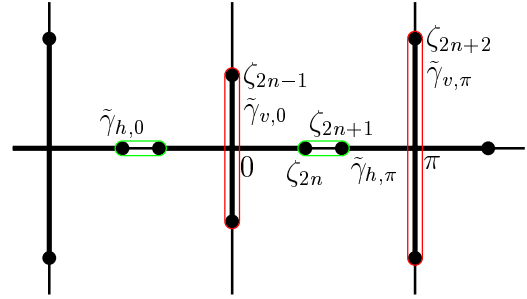}
  \label{TIBMfig:2}
  \caption{Les lacets pour les phases}
\end{floatingfigure}
%%%%%%%%%%%%%%%%%%%%%%%%%%%%%%%%%%%%%%%%%%%%%%%%%%%%%%%%%%%%%%%%%%%%% %
%
\noindent L'impulsion complexe se prolonge analytiquement le long des
lacets $\tilde\gamma_{h,0}$, $\tilde\gamma_{h,\pi}$,
$\tilde\gamma_{v,0}$ et $\tilde\gamma_{v,\pi}$ (voir la
figure~\ref{TIBMfig:2}). Donc, ces lacets sont les projections (i.e.
image par $\Pi$) sur le plan des $\zeta$ de lacets de $\Gamma$ ;
appelons-les respectivement $\gamma_{h,0}$, $\gamma_{h,\pi}$,
$\gamma_{v,0}$ et $\gamma_{v,\pi}$ (i.e.
$\tilde\gamma_{h,0}=\Pi(\gamma_{h,0})$,
$\tilde\gamma_{h,\pi}=\Pi(\gamma_{h,\pi})$, etc).  Nous les avons
repr{\'e}sent{\'e} sur la figure~\ref{TIBMfig:actions}. Le lacet
$\gamma_{h,\pi}$ connecte les branches r{\'e}elles $\gamma_\pi$ et
$\gamma_0$ ; le lacet $\gamma_{h,0}$ connecte les branches r{\'e}elles
$\gamma_0$ et $\gamma_\pi-(2\pi,0)$. Le lacet $\gamma_{v,\pi}$
connecte les branches r{\'e}elles $\gamma_\pi$ et $\gamma_\pi+(0,2\pi)$ ;
le lacet $\gamma_{v,0}$ connecte les branches r{\'e}elles $\gamma_0$ et
$\gamma_0+(0,2\pi)$.
\subsection{Les int{\'e}grales de phase, les int{\'e}grales d'action et les
  coefficients de tunnel}
\label{sec:acti-integr-tunn-1}
Soit $\sigma\in\{0,\pi\}$. Au lacet $\gamma_\sigma$, nous associons
{\it l'int{\'e}grale de phase} $\Phi_\sigma$ d{\'e}finie par
\begin{equation}
  \label{Phi:Gamma}
  \Phi_\sigma(E)=\frac12\oint_{\gamma_\sigma}\kappa\,d\zeta.
\end{equation}
La fonction $E\mapsto\Phi_\sigma(E)$ est analytique r{\'e}elle sur $J$.
Elle ne s'annule pas pour $E$ dans $J$. Le sens d'int{\'e}gration est
choisi de fa{\c c}on {\`a} ce que $\Phi_\sigma(E)$ soit positive. On montre
alors qu'il existe $c>0$ tel que
\begin{equation}
  \label{eq:21}
  \forall E\in J,\quad\Phi_0'(E)<-c\quad\text{ et
  }\quad\Phi_\pi'(E)>c.
\end{equation}
\smallpagebreak Au lacet $\gamma_{v,\sigma}$, nous associons {\it
  l'int{\'e}grale d'action verticale} $S_{v,\sigma}$ d{\'e}finie par
\begin{equation}\label{Sv:Gamma}
  S_{v,\sigma}(E)=-\frac i2\oint_{\gamma_{v,\sigma}}\kappa d\zeta,
\end{equation}
et le {\it coefficient de tunnel vertical}
\begin{equation}
  \label{tv}
  t_{v,\sigma}(E)=\exp\left(-\frac1\varepsilon S_{v,\sigma}(E)\right).
\end{equation}
La fonction $E\mapsto S_{v,\sigma}(E)$ est analytique r{\'e}elle et elle ne
s'annule pas sur $J$. Par d{\'e}finitions, le sens d'int{\'e}gration est choisi
de fa{\c c}on {\`a} ce que $S_{v,\sigma}(E)$ soit positive.
\smallpagebreak L'indice $\sigma$ {\'e}tant choisi comme ci-dessus, nous
d{\'e}finissons {\it l'int{\'e}grale d'action horizontale} $S_{h,\sigma}$ par
\begin{equation}
  \label{Sh:Gamma}
  S_{h,\sigma}(E)=-\frac i2\oint_{\gamma_{h,\sigma}} \kappa(\zeta)\,d\zeta.
\end{equation}
La fonction $E\mapsto S_{h,\sigma}(E)$ est analytique r{\'e}elle et elle
ne s'annule pas sur $J$. Par d{\'e}finition, nous choisissons
l'orientation du contour d'int{\'e}gration afin que $S_{h,\sigma}(E)$ soit
positive. Le {\it coefficient de tunnel horizontal} est d{\'e}fini par
\begin{equation}
  \label{th}
  t_{h,\sigma}(E)=\exp\left(-\frac1\varepsilon S_{h,\sigma}(E)\right).
\end{equation}
Comme le cosinus est une fonction paire, on calcule
\begin{equation}
  \label{parity-h}
  S_{h,0}(E)=S_{h,\pi}(E)\quad\text{ et
  }\quad t_{h,0}(E)=t_{h,\pi}(E).
\end{equation}
On d{\'e}finit
\begin{equation}
  \label{eq:15}
  S_h(E)=S_{h,0}(E)+S_{h,\pi}(E)\quad\text{ et }\quad
  t_h(E)=t_{h,0}(E)\cdot t_{h,\pi}(E).
\end{equation}
\section{Une famille ergodique}
\label{sec:une-famille-ergod}
Avant discuter les propri{\'e}t{\'e}s spectrales de $H_{\zeta,\varepsilon}$,
nous rappelons quelques r{\'e}sultats g{\'e}n{\'e}raux bien connus de la th{\'e}orie
spectrale des op{\'e}rateurs quasi-p{\'e}riodiques.
\smallpagebreak Comme $2\pi/\varepsilon$ est suppos{\'e} irrationnel, la
fonction $x\mapsto V(x-z)+\alpha \cos(\varepsilon x)$ est
quasi-p{\'e}riodique en $x$, et les op{\'e}rateurs d{\'e}finis par~\eqref{family}
forment une famille ergodique (voir~\cite{Pa-Fi:92}).
\smallpagebreak L'ergodicit{\'e} entra{\^\i}ne imm{\'e}diatement les cons{\'e}quences
suivantes :
\begin{enumerate}
\item le spectre de $H_{\zeta,\varepsilon}$ est presque s{\^u}rement
  ind{\'e}pendant de $\zeta$ (\cite{Pa-Fi:92}) ; en fait, dans le cas des
  op{\'e}rateurs quasi-p{\'e}riodiques, il est ind{\'e}pendant de $\zeta$
  (\cite{Av-Si:83}) ;
\item les composantes spectrales, {\`a} savoir les spectres absolument
  continu, singulier continu et purement ponctuel (c'est-{\`a}-dire
  l'adh{\'e}rence de l'ensemble des valeurs propres) sont presque s{\^u}rement
  ind{\'e}pendantes de $\zeta$ (\cite{Pa-Fi:92}) ;
\item le spectre absolument continu ne d{\'e}pend pas de $\zeta$
  (\cite{MR2000f:47060}) ;
\item le spectre discret est vide (\cite{Pa-Fi:92}) ;
\item l'exposant de Liapounoff existe pour presque tout $\zeta$ et
  n'en d{\'e}pend pas (\cite{Pa-Fi:92}) ; il est d{\'e}finit de la fa{\c c}on
  suivante : soit $x\mapsto\psi(x)$, la solution du probl{\`e}me de Cauchy
  \begin{equation*}
    H_{\zeta,\varepsilon}\psi=E\psi,\quad\psi_{|x=0}=0,\quad
    \psi'_{|x=0}=1,
  \end{equation*}
  la limite suivante existe et d{\'e}finit l'exposant de Liapounoff
  \begin{equation*}
      \Theta(E)=\Theta(E,\varepsilon):=\lim_{x\to+\infty}\frac{\log
    \left(\sqrt{|\psi(x,E,\zeta)|^2+|\psi'(x,E,\zeta)|^2}\right)}{|x|}.
  \end{equation*}
\item le spectre absolument continu est la cl{\^o}ture essentielle de
  l'ensemble des {\'e}nergies auxquelles l'exposant de Liapounoff s'annule
  (Th{\'e}or{\`e}me de Ishii-Pastur-Kotani (\cite{Pa-Fi:92}))
\item la densit{\'e} d'{\'e}tats int{\'e}gr{\'e}e existe pour presque tout $\zeta$ et
  n'en d{\'e}pend pas (\cite{Pa-Fi:92}) ; elle est d{\'e}finit de la fa{\c c}on
  suivante : pour $L>0$, soit $H_{\zeta,\varepsilon;L}$, l'op{\'e}rateur
  $H_{\zeta,\varepsilon}$ restreint {\`a} l'intervalle $[-L,L]$ avec les
  conditions de Dirichlet au bord ; pour presque tout $\zeta$, pour
  $E\in\R$, la limite suivante existe
  \begin{equation*}
    N(E)=N(E,\varepsilon):=\lim_{L\to+\infty}\frac{\#\{\text{ valeurs propres de
    }H_{\zeta,\varepsilon;L}\text{ inf{\'e}rieures {\`a} }E\}}{2L}
  \end{equation*}
  et ne d{\'e}pend pas de $\zeta$ ; c'est la densit{\'e} d'{\'e}tats int{\'e}gr{\'e}e de
  $H_{\zeta,\varepsilon}$.
\item le spectre de $H_{\zeta,\varepsilon}$ est l'ensemble des points
  de croissance stricte de la densit{\'e} d'{\'e}tats int{\'e}gr{\'e}e.
\end{enumerate}
Apr{\`e}s ces rappels de propri{\'e}t{\'e}s g{\'e}n{\'e}rales des op{\'e}rateurs de
Schr{\"o}dinger quasi-p{\'e}riodiques g{\'e}n{\'e}rales, revenons au
mod{\`e}le~\eqref{family}.
\section{Une description grossi{\`e}re du lieu du spectre dans $J$}
\label{sec:une-descr-gross}
Nous supposerons dor{\'e}navant, et ce dans toute la suite du texte, que les
hypoth{\`e}ses (H), (O) sont satisfaites et que $J$ est un intervalle compact
v{\'e}rifiant (BEI).  Nous supposerons de plus que,
\begin{description}
\item[(T) ] pour $E\in J$,
  \begin{equation*}
    2\pi\cdot\min(\im\zeta_{2n-2}(E),\,\im\zeta_{2n+3}(E))
    >\max(S_h(E),\,S_{v,0}(E),\,S_{v,\pi}(E)).
  \end{equation*}
\end{description}
Cette hypoth{\`e}se est n{\'e}cessaire pour obtenir les r{\'e}sultats les plus
simples (voir~\cite{Fe-Kl:04a} pour plus de d{\'e}tails). Elle est
v{\'e}rifi{\'e}e s'il y a deux bandes cons{\'e}cutives assez proches l'une de
l'autre et suffisamment {\'e}loign{\'e}es du reste du spectre. Remarquons que
les r{\'e}sultats de cet article restent vrais sous une hypoth{\`e}se plus
faible (mais moins explicite) que~(T). Lorsque~(T) n'est pas v{\'e}rifi{\'e}e,
d'autres coefficients d'effet tunnel que ceux pr{\'e}sent{\'e}s dans la
partie~\ref{sec:iso-energy-curve-3} peuvent entrer en jeu.
\smallpagebreak Soit
  \begin{equation}
    \label{eq:6}
    \delta_0=\frac12\inf_{E\in J}\min(S_h(E),S_{v,0}(E),S_{v,\pi}(E))>0.
  \end{equation}
\smallpagebreak On d{\'e}montre
\begin{Th}
  \label{thr:2} 
  Fixons $E_*\in J$. Pour $\varepsilon$ suffisamment petit, il existe
  $V_*\subset \C$, un voisinage de $E_*$ ind{\'e}pendant de $\varepsilon$,
  et deux fonctions analytiques r{\'e}elles $E\mapsto\check
  \Phi_0(E,\varepsilon)$ et $E\mapsto\check\Phi_\pi(E,\varepsilon)$,
  d{\'e}finies dans $V_*$ et v{\'e}rifiant
  \begin{equation}
    \label{eq:17}
    \check\Phi_0(E,\varepsilon)=\Phi_0(E)+o(\varepsilon),\quad
    \check\Phi_\pi(E,\varepsilon)=\Phi_\pi(E)+o(\varepsilon)\quad\text{quand
    }\varepsilon\to0,
  \end{equation}
  telles que, si on d{\'e}finit les deux suites finies de points de $J\cap
  V_*$, $(E_{0}^{(l)})_l:=(E_{0}^{(l)}(\varepsilon))_l$ et
  $(E_{\pi}^{(l')})_{l'}:=(E_{\pi}^{(l')}(\varepsilon))_{l'}$ par
  \begin{equation}
    \label{eq:11}
    \frac1{\varepsilon}\check\Phi_0(E_0^{(l)},\varepsilon)=\frac\pi2+\pi
    l\quad\text{ et }\quad\frac1{\varepsilon}\check
    \Phi_\pi(E_\pi^{(l')},\varepsilon)=\frac\pi2+\pi l',\quad
    (l,\,l')\in\N^2,
  \end{equation}
  alors, pour $\varepsilon$ suffisamment petit, pour tout $\zeta$
  r{\'e}el, le spectre de $H_{\zeta,\varepsilon}$ dans $J\cap V_*$ est
  contenu dans la r{\'e}union des intervalles
  \begin{equation}
    \label{eq:16}
    I_0^{(l)}:=
    E_{0}^{(l)}+[-e^{-\delta_0/\varepsilon},e^{-\delta_0/\varepsilon}]
    \quad\text{ et }\quad
    I_{\pi}^{(l')}:=E_{\pi}^{(l')}+[-e^{-\delta_0/\varepsilon},
    e^{-\delta_0/\varepsilon}]
  \end{equation}
  c'est-{\`a}-dire
  \begin{equation}
    \label{eq:5}
    \sigma(H_{\zeta,\varepsilon})\cap J\cap V_*\subset\left(\bigcup_{l}
    I_{0}^{(l)}\right)\bigcup\left(\bigcup_{l'}
    I_{\pi}^{(l')}\right).
  \end{equation}
\end{Th}
\noindent Dans la suite pour simplifier les notations, on omettra la
r{\'e}f{\'e}rence {\`a} $\varepsilon$ dans les fonctions $\check\Phi_0$ et
$\check\Phi_\pi$.
\smallpagebreak D'apr{\`e}s~\eqref{eq:21} et~\eqref{eq:17}, il existe
$C>0$ telle que, pour $\varepsilon$ suffisamment petit, les points
d{\'e}finis en~\eqref{eq:11} v{\'e}rifient
\begin{gather}
  \label{eq:7}
  \frac{1}{C}\varepsilon\le E_0^{(l)}-E_0^{(l-1)}\le
  C\varepsilon,\quad l=L_{0}^-+1,\dots,L_{0}^+\\
  \label{eq:8}
  \frac{1}{C}\varepsilon\le E_{\pi}^{(l)}-E_{\pi}^{(l-1)}\le
  C\varepsilon,\quad l=L_{\pi}^-+1,\dots,L_{\pi}^+.
\end{gather}
De plus, pour $\nu\in\{0,\pi\}$, dans l'intervalle $J\cap V_*$, le
nombre de points $E_{\nu}^{(l)}$ est d'ordre $1/\varepsilon$.
\smallpagebreak On dira dans la suite que les points $E_{0}^{(l)}$
(resp. $E_{\pi}^{(l)}$), et, par extension, les intervalles
$I_{0}^{(l)}$ (resp. $I_{\pi}^{(l)}$) qui y sont attach{\'e}s, sont de
type $0$ (resp. de type $\pi$).
\smallpagebreak Par~\eqref{eq:7} et~\eqref{eq:8}, les intervalles de
type $0$ (resp. $\pi$) sont disjoints et, tout intervalle de type $0$
(resp. $\pi$) rencontre au plus un intervalle de type $\pi$ (resp.
$0$).
\section{La description pr{\'e}cis{\'e}e du spectre dans $J$}
\label{sec:la-descr-prec}
On va maintenant d{\'e}crire le spectre dans l'un des intervalles d{\'e}finis
dans le Th{\'e}o\-r{\`e}me~\ref{thr:2}. Pour ce faire, il va falloir
distinguer deux cas selon que cet intervalle, supposons le de type
$0$, rencontre ou non un intervalle de type $\pi$. Les intervalles de
l'une des famille qui ne coupent aucun intervalle de l'autre famille
seront dit {\it non-r{\'e}sonant}, les autres {\'e}tant les intervalles {\it
  r{\'e}sonants}. Comme dit au paragraphe pr{\'e}c{\'e}dent, {\`a} chaque intervalle
r{\'e}sonant de l'une des familles (s'il en existe) correspond un unique
intervalle r{\'e}sonant de l'autre famille. Quand nous discuterons les
intervalles r{\'e}sonants, nous d{\'e}crirons le spectre dans la r{\'e}union des
deux intervalles.
\smallpagebreak On peut se demander si des {\'e}nergies r{\'e}sonantes
existent.  Comme la d{\'e}riv{\'e}e de $\Phi_\pi$ (resp. $\Phi_0$) est
strictement positive (resp. n{\'e}gative) sur $J$, alors, pour
$\varepsilon$ suffisamment petit, il en est de m{\^e}me pour celle de
$\check\Phi_\pi$ (resp. $\check\Phi_0$). Donc, approximativement,
lorsque $\varepsilon$ d{\'e}croit vers $0$, les points de type $\pi$ se
d{\'e}placent vers la gauche et ceux de la suite de type $0$ vers la
droite; comme les points ont un mouvement continu en $\varepsilon$,
ils se rencontrent. Ainsi, quitte {\`a} r{\'e}duire $\varepsilon$, on peut
donc toujours cr{\'e}er des intervalles r{\'e}sonants ! Clairement, ceci peut
{\^e}tre r{\'e}alis{\'e} dans tout sous-intervalle relativement compact de
l'int{\'e}rieur de $J\cap V_*$ de longueur au moins $C\varepsilon$ si $C$
est suffisamment grand.
\begin{Rem}
  \label{rem:2}
  Pour une fonction $V$ g{\'e}n{\'e}rique, il n'y a que peu d'intervalles
  r{\'e}sonants. Mais, si $V$ pr{\'e}sente des sym{\'e}tries particuli{\`e}res, par
  exemple si $V$ est paire, alors tous les points $E_0^{(l)}$ et
  $E_\pi^{(l')}$ co{\"\i}ncident et tous les intervalles sont r{\'e}sonants !
  Ceci est d{\^u} {\`a} la parit{\'e} du cosinus ; ce n'est plus vrai en g{\'e}n{\'e}ral
  si on remplace le potentiel $\alpha\cos(\cdot)$ par un autre
  potentiel.
\end{Rem}
\subsection{Le cas des intervalles non-r{\'e}sonants}
\label{sec:le-cas-des}
C'est le cas le plus simple. On d{\'e}crira les r{\'e}sultats dans le cas de
la famille $\pi$ ; la transposition {\`a} la famille $0$ est imm{\'e}diate.
\begin{Th}
  \label{th:tib:sp:1}
  Pla{\c c}ons nous dans les conditions du Th{\'e}or{\`e}me~\ref{thr:2}. Pour
  $\varepsilon$ suffisamment petit, soient $(I_0^{(l')})_{l'}$ et
  $(I_\pi^{(l)})_{l}$, les suites finies d'intervalles d{\'e}finies dans
  ce r{\'e}sultat. Consid{\'e}rons $l$ tel que, pour tout $l'$,
  $I_\pi^{(l)}\cap I_0^{(l')}=\emptyset$. Alors, il existe $\check
  E_\pi^{(l)}$ et $\check w_\pi^{(l)}$ v{\'e}rifiant
  \begin{gather}
    \label{eq:24}
    \check E_\pi^{(l)}=E_\pi^{(l)}+\varepsilon\,\frac{\Lambda_n(V)}{2
      \check\Phi_\pi'(E_\pi^{(l)})}\,t_h(E_\pi^{(l)})\,
    \tan\left(\frac{\check\Phi_0(E_\pi^{(l)})}{\varepsilon}\right)
    \,(1+o(1)),\\
    \label{eq:19} \check w_\pi^{(l)}=
    \frac{\varepsilon}{\check\Phi_\pi'(E_\pi^{(l)})}
    \left(\frac{t_h(E_\pi^{(l)})}
      {2\left|\cos\left(\frac{\check\Phi_0(E_\pi^{(l)})}
            {\varepsilon}\right)\right|}+
      t_{v,\pi}(E_\pi^{(l)})\right)\,(1+o(1)),
  \end{gather}
  tels que, si on d{\'e}finit
  \begin{equation}
    \label{checkI}
    \check I_\pi^{(l)}:= [\check E_\pi^{(l)}-\check w_\pi^{(l)},\check
    E_\pi^{(l)}+\check w_\pi^{(l)}]
  \end{equation}
  alors
  \begin{equation*}
    \sigma(H_{\zeta,\varepsilon})\cap I_\pi^{(l)}\subset\check I_\pi^{(l)}
  \end{equation*}
  De plus, si $dN_{\varepsilon}(E)$ est la mesure de densit{\'e} d'{\'e}tats
  $H_{\zeta,\varepsilon}$, alors
  \begin{equation*}
    \int_{\check I_\pi^{(l)}}dN_{\varepsilon}(E)=\frac\varepsilon{2\pi}.
  \end{equation*}
\end{Th}
\noindent On remarque que, dans le cas non-r{\'e}sonant, pour chacune des
famille d'intervalles, la description du lieu du spectre est semblable
{\`a} celle obtenue pour le fond de spectre dans le
travail~\cite{MR2003f:82043}.  N{\'e}anmoins, m{\^e}me dans le cas
non-r{\'e}sonant, on constate l'influence de l'une des suites sur l'autre
dans les formules~\eqref{eq:24} et~\eqref{eq:19}. Soit $E_0$, le point
de la suite $E_0^{(l')}$ le plus proche {\`a} $E_\pi:=E_\pi^{(l)}$.
Consid{\'e}rons le deuxi{\`e}me terme dans la formule~\eqref{eq:24} d{\'e}crivant
le centre de l'intervalle $\check I_\pi^{(l)}$. Comme $\check
\Phi_\pi'(E)>0$, le signe de ce terme est celui du facteur
$\tan\left(\frac{\check\Phi_0(E_\pi)}{\varepsilon}\right)$. Supposons
que $E_0$ et $E_\pi$ sont assez proches l'un de l'autre.  Comme
$\frac1\varepsilon\check\Phi_0(E_0)=\frac\pi2\,\text{mod}\,\pi$ par
d{\'e}finition et comme $\check \Phi_0'(E)<0$, le deuxi{\`e}me terme
de~\eqref{eq:24} est n{\'e}gatif si $E_\pi$ est {\`a} gauche de $E_0$, et
positif si $E_\pi$ est {\`a} droite de $E_0$. Donc, on voit qu'il y a une
r{\'e}pulsion entre les intervalles $\check I_0$ et $\check I_\pi$. Comme
la distance du point $E_\pi$ {\`a} $E_0$ est contr{\^o}l{\'e}e par le facteur
\begin{equation}
  \label{eq:25}
  \cos\left(\frac{\check\Phi_0(E_\pi^{(l)})}{\varepsilon}\right),
\end{equation}
la r{\'e}pulsion est d'autant plus importante que $|E_0-E_\pi|$ est petit.\\
\smallpagebreak Discutons maintenant la nature du spectre dans l'intervalle
$\check I_\pi^{(l)}$. Soit
\begin{equation}
  \label{lambda_pi}
  \lambda_\pi(E)=
  \frac{t_{v,\pi}(E)}{t_h(E)}\,\dist\left(E,\bigcup_{l}\{E_0^{(l')}\}\right),
\end{equation}
o{\`u}, pour $A$ un ensemble d'{\'e}nergies, $\dist(E,A)$ d{\'e}signe la distance
de $E$ {\`a} l'ensemble $A$. On d{\'e}montre le
\begin{Th}
  \label{th:gamma-pi}
  Sur l'intervalle $\check I_\pi^{(l)}$, l'exposant de Liapounoff
  admet l'asymptotique
  \begin{equation}
    \label{gamma-pi}
    \Theta(E,\varepsilon)=\frac\varepsilon{2\pi}
    \log^+\lambda_\pi(E_\pi^{(l)})+o(1),
  \end{equation}
  o{\`u} $o(1)$ tend vers $0$ quand $\varepsilon$ tends vers $0$. Ici,
  $\log^+$ d{\'e}signe la partie positive du logarithme naturel, i.e.
  $\log^+=\max(0,\log)$.
\end{Th}
\noindent Supposons que $(S_h-S_{v,\pi})(E_\pi^{(l)})>0$. Si
$\dist\left(E_\pi^{(l)},\bigcup_{l}\{E_0^{(l')}\}\right)
\geq\varepsilon ^N$ (o{\`u} $N$ {\'e}tant un entier positif fix{\'e}) alors le
Th{\'e}or{\`e}me~\ref{th:gamma-pi} et la formule~\eqref{lambda_pi} impliquent
que
\begin{equation}
  \label{eq:10}
  \Theta(E,\varepsilon)=\frac1{2\pi}(S_h-S_{v,\pi})
  (E_\pi^{(l)})+o(1)\text{ quand }\varepsilon\to0.
\end{equation}
Par contre, quand $E_\pi^{(l)}$ est seulement {\`a} une distance
$e^{-\delta/\varepsilon}$ (pour $0<\delta<(S_h-S_{v,\pi})^+$) de
l'ensemble des points $E_0^{(l')}$, sur $\check I_\pi^{(l)}$, on a
\begin{equation*}
 \Theta(E,\varepsilon)=\frac1{2\pi}(S_h-S_{v,\pi})
  (E_\pi^{(l)})-\delta+o(1)\text{ quand }\varepsilon\to0.
\end{equation*}
On voit donc que la valeur de $\Theta$ sur $\check I_\pi^{(l)}$ chute
brutalement lorsque l'on approche $E_\pi^{(l)}$ de la suite de points
$(E_0^{(l')})_{l'}$.\\
Le Th{\'e}or{\`e}me~\eqref{th:gamma-pi} entra{\^\i}ne le
\begin{Cor}
  \label{cor:tib:sp:3}
  Fixons $c>0$. Pour $\varepsilon$ suffisamment petit, si
  $I_\pi^{(l)}$ est non-r{\'e}sonant et si $\varepsilon \log
  \lambda_\pi(E_\pi^{(l)})>c$, alors, l'intervalle $\check
  I_\pi^{(l)}$ d{\'e}fini dans le Th{\'e}or{\`e}me~\ref{th:tib:sp:1} ne contient
  que du spectre singulier.
\end{Cor}
\noindent Si la quantit{\'e} $\lambda_\pi$ est petite sur l'intervalle
$\check I_\pi^{(l)}$, la plus grande partie de cet intervalle est dans
le spectre absolument continue ; on d{\'e}montre le
\begin{Th}
  \label{th:tib:sp:2}
  Pour $c>0$, il existe $\eta$, une constante positive, et un ensemble
  de nombres diophantiens $D\subset (0,1)$ tels que
  \begin{itemize}
  \item $D$ est asymptotiquement de mesure totale i.e.
    \begin{equation*}
      \frac{\mes(D\cap(0,\varepsilon))}{\varepsilon}=
      1+o\left(e^{-\eta/\varepsilon}\right)\text{
        lorsque }\varepsilon\to0.
    \end{equation*}
  \item pour $\varepsilon\in D$ suffisamment petit, si $\check
    I_\pi^{(l)}$ est non r{\'e}sonant et si $\varepsilon\log
    \lambda_\pi(E_\pi^{(l)})<-c$, alors l'intervalle $\check
    I_\pi^{(l)}$ d{\'e}fini dans le Th{\'e}or{\`e}me~\ref{th:tib:sp:1} contient
    majoritairement du spectre absolument continu de
    $H_{\zeta,\varepsilon}$ ; plus pr{\'e}cis{\'e}ment, on a
    \begin{equation*}
      \frac{\mes(\check I_\pi^{(l)}\cap \Sigma_{\rm ac})}
        {\mes(\check I_\pi^{(l)})}=1+o(1),
    \end{equation*}
    o{\`u}, $\Sigma_{ac}$ d{\'e}signe le spectre absolument continu de
    $H_{\zeta,\varepsilon}$.
  \end{itemize}
\end{Th}
\noindent Pour les intervalles de la famille de type $0$, on obtient
bien s{\^u}r les pendants des
Th{\'e}or{\`e}\-mes~\ref{th:tib:sp:1},~\ref{th:gamma-pi},~\ref{th:tib:sp:2} et
du Corollaire~\ref{cor:tib:sp:3}.
\smallpagebreak Ces r{\'e}sultats appellent quelques commentaires.
\smallpagebreak Comme nous l'avons vu {\`a} la fin de la
section~\ref{sec:une-descr-gross}, en choisissant judicieusement
$\varepsilon$, on peut s'arranger pour que la distance minimale entre
les points de la suite $0$ et ceux de la suite $\pi$ soit aussi petite
que souhait{\'e}e ; de plus, ceci peut {\^e}tre fait dans tout sous-intervalle
relativement compact de l'int{\'e}rieur de $J$ de longueur au moins
$C\varepsilon$ si $C$ est suffisamment grand. Or, sur un tel
intervalle, les actions $E\mapsto S_h(E)$, $E\mapsto S_{v,0}(E)$ et
$E\mapsto S_{v,\pi}(E)$ varient d'au plus $C'\varepsilon$. Donc,
quitte {\`a} choisir $\varepsilon$ assez petit correctement, on peut
essentiellement supposer qu'il existe un point de la suite de type $0$
et un point de la suite de type $\pi$ s{\'e}par{\'e}s d'une distance
arbitraire, inf{\'e}rieure {\`a} $\varepsilon$ tels que, sur un voisinage de
taille $\varepsilon$ de ces points, le tripl{\'e} des actions
$E\mapsto(S_h(E),S_{v,0}(E),S_{v,\pi}(E))$ prenne l'une quelconque de
ses valeurs possibles sur $J$. On peut donc choisir les grandeurs
$E_\pi^{(l')}-E_0^{(l)}$ et $(S_h(E),S_{v,0}(E),S_{v,\pi}(E))$
essentiellement ind{\'e}pendamment l'une de l'autre.
\smallpagebreak Dans le cas o{\`u} (BEI) est v{\'e}rifi{\'e}e, deux nouveaux
ph{\'e}nom{\`e}nes spectraux peuvent appara{\^\i}tre. Nous allons les d{\'e}crire
maintenant.
\subsubsection{Transitions dues {\`a} l'approche de la situation
  r{\'e}sonante}
\label{sec:trans-dues-lappr}
On voit que la nature du spectre sur les intervalles d{\'e}finis dans le
Th{\'e}or{\`e}me~\ref{th:tib:sp:1} d{\'e}pend de leur proximit{\'e} avec les
intervalles de l'autre famille. L'effet de l'interaction entre les
deux familles d'intervalles peut aller jusqu'{\`a} entra{\^\i}ner des
changement dans la nature du spectre par rapport {\`a} ce qu'on voit
lorsque la distance entre les points de deux familles reste d'ordre
$\varepsilon$. Prenons un exemple. Supposons que l'intervalle $J$
v{\'e}rifie les hypoth{\`e}ses:
\begin{gather}
 \label{eq:33}
  \min_{E\in J} S_h(E)>\max_{\nu\in\{0,\pi\}}\max_{E\in
  J}S_{v,\nu}(E),\\
  \intertext{et}
  \label{eq:33a}
 \frac32  \min_{\nu\in\{0,\pi\}}\min_{E\in J}S_{v,\nu}(E)>\max_{E\in J} S_h(E).
\end{gather}
Sous la condition~\eqref{eq:33}, on obtient
$\D\delta_0=\frac12\,\min_{\nu\in\{0,\pi\}}\min_{E\in J}S_{v,\nu}(E)$.
Donc, il existe $c>0$ tel que, pour $E\in J$,
\begin{equation}
  \label{eq:33b}
    S_h(E)-S_{v,\nu}(E)-\delta_0<-c<0, \quad \nu=0,\pi.
\end{equation}
Consid{\'e}rons deux intervalles $I_0^{(l')}$ et $I_\pi^{(l)}$
non r{\'e}sonants situ{\'e}s dans $J\cap V_*$. On a 
\begin{itemize}
\item si ces deux intervalles sont s{\'e}par{\'e}s par une distance de taille
  $\varepsilon^N$ (o{\`u} $N$ est un entier fix{\'e}), la
  relation~\eqref{eq:33} garantit que, sur ces intervalles, le spectre
  est contr{\^o}l{\'e} par le Corollaire~\ref{cor:tib:sp:3} et son pendant
  pour la suite de type $0$.
\item si ces deux intervalles sont s{\'e}par{\'e}s par une distance d'ordre
  $\exp\left(-\delta_0/\varepsilon\right)$, alors la
  relation~\eqref{eq:33b} garantit que, sur ces intervalles, le
  spectre est contr{\^o}l{\'e} par le Th{\'e}or{\`e}me~\ref{th:tib:sp:2} et son
  pendant pour la suite de type
  $0$.
\end{itemize}
\vskip.1cm L'existence d'intervalles $J$ pour lesquels on
a~\eqref{eq:33} et~\eqref{eq:33a} peut {\^e}tre v{\'e}rifi{\'e}e num{\'e}rique\-ment,
voir la partie~\ref{sec:phase-diagram}. Ainsi, non seulement
l'emplacement du spectre d{\'e}pend de la distance s{\'e}parant localement les
suites de type $0$ et de type $\pi$, mais la nature du spectre en
d{\'e}pend {\'e}galement.  Il peut s'op{\'e}rer une transition: du spectre qui
serait singulier lorsque les deux suites sont {\'e}loign{\'e}es l'une de
l'autre peut devenir absolument continu lorsqu'elles sont proches
(voir la figure~\ref{fig:res_tun}). Notons que la transition ne peut
s'op{\'e}rer dans l'autre sens : le spectre absolument continu lorsque les
suites sont {\'e}loign{\'e}es le reste si elles sont proches.
\subsubsection{Alternance des types spectraux}
\label{sec:alternance-des-types}
\begin{figure}[htbp]
  \centering \subfigure[Transitions r{\'e}sonantes]{
    \includegraphics[bbllx=71,bblly=676,bburx=218,bbury=721,width=6cm]{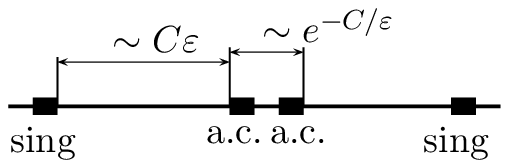}
  \label{fig:res_tun}}
\hskip3cm \subfigure[Alternance spectrale]{
  \includegraphics[bbllx=71,bblly=676,bburx=218,bbury=721,width=6cm]{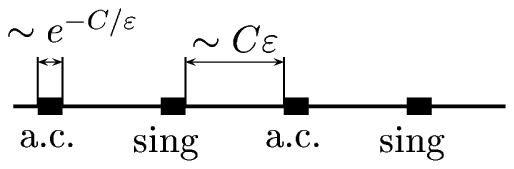}
  \label{fig:alt_spec}}
  \caption{Deux nouveaux ph{\'e}nom{\`e}nes spectraux d{\^u}s {\`a} des r{\'e}sonances faibles}
  \label{fig:contours}
\end{figure}

De plus, un autre nouveau ph{\'e}nom{\`e}ne spectral peut appara{\^\i}tre. Pour
simplifier, supposons que, dans $V_*\cap J$, les distances entre les
points $\{E_0^{(l)}\}$ et les points $\{E_\pi^{(l')}\}$ sont
sup{\'e}rieures {\`a} $\varepsilon^N$, o{\`u} $N$ est un entier fix{\'e} (c'est-{\`a}-dire
que toutes les {\'e}nergies sont non r{\'e}sonantes). En tenant compte du
Th{\'e}or{\`e}me~\ref{th:tib:sp:2} et du Corollaire~\ref{cor:tib:sp:3}, ceci
implique que, sur $\check I_0^{(l)}$ (resp. $\check I_\pi^{(l')}$), la
nature du spectre est d{\'e}termin{\'e}e par la taille des quotients $t_{v,0}
(E_0^{(l)})/t_{h}(E_0^{(l)})$ (resp.  $t_{v,\pi}(E_\pi^{(l')})
/t_{h}(E_\pi^{(l')})$) par rapport {\`a} $1$.  S'il existe $\delta>0$ tel
que
\begin{equation}
  \label{eq:22}
  \forall E\in J_*,\quad S_{v,\pi}(E)-S_h(E)>\delta\quad\text{ et
  }\quad S_{v,0}(E)-S_h(E)<-\delta,
\end{equation}
alors, dans $V_*\cap J$, les suites de type $0$ et de type $\pi$
contiennent du spectre de types spectraux \og oppos{\'e}s \fg, i.e. le spectre
dans les intervalles de type $0$ est singulier, et celui dans les
intervalles de type $\pi$ est, pour l'essentiel, absolument continu.
Ceci ne vaut bien s{\^u}r que si $\varepsilon$ est suffisamment petit et
v{\'e}rifie la condition diophantienne impos{\'e}e dans le
Th{\'e}or{\`e}me~\ref{th:tib:sp:2} On obtient ainsi un entrelacement
d'ensembles de spectres de type \og oppos{\'e}s \fg, voir la
figure~\ref{fig:alt_spec}. Dans ce cas, le nombre de transitions
d'Anderson dans l'intervalle $V_*\cap J$ est de l'ordre de
$1/\varepsilon$.\\
{\`A} l'instar des conditions~\eqref{eq:33} et~\eqref{eq:33a}, on v{\'e}rifie
num{\'e}riquement que la condition~\eqref{eq:22} est effectivement
r{\'e}alisable (voir section~\ref{sec:phase-diagram}).
%
%%%%%%%%%%%%%%%%%%%%%%%%%%%%%%%%%%%%%%%%%%%%%%%%%%%%%%%%%%%%%%%%%%%%%% %
%%%%%%%%%%%%%%%%%%%%%%%%%%%%%%%%%%%%%%%%%%%%%%%%%%%%%%%%%%%%%%%%%%%%%% %
%
\subsection{Les r{\'e}sonances}
\label{sec:plus-proche-de}
On consid{\`e}re maintenant deux {\'e}nergies, $E_0^{(l)}$ et $E_\pi^{(l')}$,
r{\'e}sonantes, c'est-{\`a}-dire telles que
\begin{equation}
  \label{eq:20}
  |E_\pi^{(l')}-E_0^{(l)}|<2e^{-\frac{\delta_0}{\varepsilon}}.
\end{equation}
Si~\eqref{eq:20} est v{\'e}rifi{\'e}e, les intervalles $I_0^{(l)}$ et
$I_\pi^{(l')}$ se rencontrent. On ne d{\'e}crira donc plus le spectre dans
chacun des intervalles s{\'e}par{\'e}ment mais dans la r{\'e}union des deux
intervalles.\\
Pour simplifier l'expos{\'e}, nous renommons les {\'e}nergies et les
intervalles r{\'e}sonants
\begin{equation}
  \label{eq:13}
  E_0:=E_0^{(l)},\quad\quad
  E_\pi:=E_\pi^{(l')},\quad\quad I_0:=I_0^{(l)},\quad\text{ et }\quad
  I_\pi:=I_\pi^{(l')}.
\end{equation}
Dans le cas r{\'e}sonant, le lieu du spectre et sa nature d{\'e}pendent de la
valeur du param{\`e}tre
\begin{equation}
  \label{tau}
  \tau=2\sqrt{\frac{t_{v,0}(\bar E)\,t_{v,\pi}(\bar E)}{t_h(\bar
      E)}}\quad\text{o{\`u}}\quad\bar E=\frac{E_\pi+E_0}2.
\end{equation}
Nous allons analyser en d{\'e}tail les cas o{\`u} $\tau\gg1$ et o{\`u} $\tau\ll1$.
Plus pr{\'e}cis{\'e}ment, nous fixons $\delta>0$ arbitraire et supposerons
soit que
\begin{gather}
  \label{eq:28}
  \forall E\in V_*\cap J,\quad
  S_h(E)-S_{v,0}(E)-S_{v,\pi}(E)\geq\delta,\quad
  (\text{qui sera le cas }\tau\gg1),\\
  \intertext{soit que}
  \label{eq:14}
   \forall E\in V_*\cap J,\quad S_h(E)-S_{v,0}(E)-S_{v,\pi}(E)\le
   -\delta,\quad (\text{qui sera le cas }\tau\ll1).
\end{gather}
Le cas o{\`u} $\tau\asymp1$ est le cas le plus compliqu{\'e} (mais aussi le
moins fr{\'e}quent). Nous n'en parlerons que bri{\`e}vement.
\smallpagebreak Les r{\'e}sultats sont d{\'e}crits avec les \og variables
locales \fg
\begin{equation}
  \label{xi}
  \xi_\nu(E)=\frac{\check \Phi'(\bar E)}\varepsilon\cdot \frac{E-
    E_\nu}{t_{v,\nu}(\bar E)}\quad\text{o{\`u}}\quad\nu\in\{0,\pi\}.
\end{equation}
\subsubsection{Quand $\tau$ est grand}
\label{sec:quand-tau-g}
On suppose maintenant que $\tau\gg1$. Dans ce cas, le lieu du spectre
pr{\`e}s des {\'e}nergies $E_0$ et $E_\pi$ est d{\'e}crit par le
\begin{Th}
  \label{th:tib-res:sp:1}
  Pla{\c c}ons nous dans les conditions du Th{\'e}or{\`e}me~\ref{thr:2}. Supposons
  que l'hypoth{\`e}se~\eqref{eq:28} est v{\'e}rifi{\'e}e. Alors, il existe
  $\varepsilon_0>0$ tel que, pour $0<\varepsilon<\varepsilon_0$, et
  des {\'e}nergies r{\'e}sonantes, $E_0$ et $E_\pi$, dans $V_*\cap J$, le
  spectre de $H_{\zeta,\varepsilon}$ dans $I_0\cup I_\pi$ est situ{\'e}
  dans deux intervalles $\check I_0$ et $\check I_\pi$ d{\'e}finis par
\begin{equation}
  \check I_0= \{E\in I_0:\ |\xi_0(E)|\leq1+o(1)\}
  \quad\text{et}\quad
  \check I_\pi= \{E\in I_\pi:\ |\xi_\pi(E)|\leq1+o(1)\}.
\end{equation}
Ici, $o(1)$ tend vers $0$ quand $\varepsilon$ tend vers $0$ et ne
d{\'e}pend que de $\varepsilon$.
Si $dN_{\varepsilon}(E)$ est la mesure de densit{\'e} d'{\'e}tats
$H_{\zeta,\varepsilon}$, alors
\begin{equation}
  \label{eq:2}
  \int_{\check I_0}dN_{\varepsilon}(E)=\int_{\check
    I_\pi}dN_{\varepsilon}(E) =\frac\varepsilon{2\pi}\
  \text{si}\ \check I_0\cap\check I_\pi=\emptyset,\quad\text{et}
  \quad\int_{\check I_0\cup \check I_\pi}dN_{\varepsilon}(E)=
  \frac\varepsilon{\pi}\ \text{sinon}.
\end{equation}
De plus, l'exposant de Liapounoff sur $\check I_0\cup\check I_\pi$ est
donn{\'e} par
\begin{equation}
  \label{Theta-res}
  \Theta(E,\varepsilon)=\frac
  \varepsilon{\pi}\log\left(\tau\sqrt{1+|\xi_0(E)|+|\xi_\pi(E)|})\right)+o(1),
\end{equation}
o{\`u} $o(1)$ tend vers $0$ quand $\varepsilon$ tend vers $0$ et ne d{\'e}pend
que de $\varepsilon$.
\end{Th}
\noindent L'{\'e}quation~\eqref{eq:2} nous dit que, si $\check I_0$ et
$\check I_\pi$ sont disjoints, alors ils contiennent tous les deux du
spectre ; s'ils ne sont pas disjoints, on sait seulement que leur
r{\'e}union contient du spectre.
\smallpagebreak Analysons plus avant les r{\'e}sultats du
Th{\'e}or{\`e}me~\ref{th:tib-res:sp:1}.
%
%%%%%%%%%%%%%%%%%%%%%%%%%%%%%%%%%%%%%%%%%%%%%%%%%%%%%%%%%%%%%%%%%%%%%% %
\begin{figure}[htbp]
  \centering
  \subfigure[$|E_\pi-E_0|\gg\max(t_{v,\pi},t_{v,0})$]{
    \includegraphics[bbllx=71,bblly=655,bburx=218,bbury=721,width=6cm]{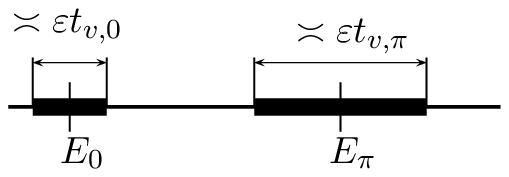}
    \label{fig:tau_grand1}}
  \hskip1cm
  \subfigure[$|E_\pi-E_0|\ll\max(t_{v,\pi},t_{v,0})$]{
    \includegraphics[bbllx=71,bblly=655,bburx=218,bbury=721,width=6cm]{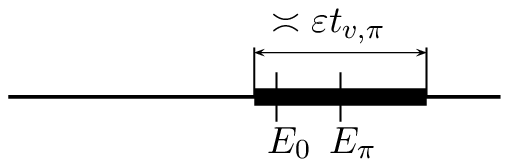}
    \label{fig:tau_grand2}}
  \caption{Le lieu du spectre quand $\tau$ est grand}
  \label{fig:tau_grand}
\end{figure}
%%%%%%%%%%%%%%%%%%%%%%%%%%%%%%%%%%%%%%%%%%%%%%%%%%%%%%%%%%%%%%%%%%%%%% %
%
\smallpagebreak{\it Le lieu du spectre.\/} Par~\eqref{xi}, les
intervalles $\check I_0$ et $\check I_\pi$ d{\'e}finis dans
Th{\'e}or{\`e}me~\ref{th:tib-res:sp:1} sont respectivement \og centr{\'e}s \fg\ aux
points $E_0$ et $E_\pi$. Leur longueurs sont donn{\'e}es par les formules:
\begin{equation*}
  |\check I_0|=\frac{2\,\varepsilon}{|\check\Phi'_0(E_0)|}\cdot
  {t_{v,0}(E_0)}\cdot(1+o(1))\quad\text{et}\quad
  |\check I_\pi|=\frac{2\,\varepsilon}{|\check\Phi'_\pi(E_\pi)|}\cdot
  {t_{v,\pi}(E_\pi)}\cdot(1+o(1)).
\end{equation*}
o{\`u} $o(1)$ tend vers $0$ quand $\varepsilon$ tend vers $0$ et ne d{\'e}pend
que de $\varepsilon$. Suivant la distance s{\'e}parant $E_\pi$ de $E_0$,
on obtient les dessins de la figure~\ref{fig:tau_grand} pour le lieu
du spectre.
\\ L'effet de r{\'e}pulsion observ{\'e} dans la section~\ref{sec:le-cas-des}
ne se manifeste pas : il est n{\'e}gligeable par rapport aux longueurs des
intervalles $\check I_0$ et $\check I_\pi$.
\smallpagebreak{\it La nature du spectre.\/} Dans les intervalles
$\check I_0$ et $\check I_\pi$, d'apr{\`e}s l'hypoth{\`e}se~\eqref{eq:28} et la
formule~\eqref{Theta-res}, l'exposant de Liapounoff est positif. Donc,
par le Th{\'e}or{\`e}me de Ishii, Pastur et Kotani (\cite{Pa-Fi:92}), dans les
deux intervalles $\check I_0$ et $\check I_\pi$, le spectre est singulier.
\smallpagebreak{\it La variation de $\Theta(E,\varepsilon)$ sur le
  spectre.\/} La formule g{\'e}n{\'e}rale~\eqref{Theta-res} peut {\^e}tre
simplifi{\'e}e de la fa{\c c}on suivante :
\begin{gather*}
  \Theta(E,\varepsilon)=\frac
  \varepsilon{\pi}\log\left(\tau\sqrt{1+|\xi_0(E)|}\right)+o(1)\quad {\rm si }
  \quad E\in \check I_\pi,\\
  \intertext{et}
  \Theta(E,\varepsilon)=\frac
  \varepsilon{\pi}\log\left(\tau\sqrt{1+|\xi_\pi(E)|}\right)+o(1)\quad {\rm si }
  \quad E\in \check I_0.
\end{gather*}
Si $|E_\pi-E_0|\gg\max(t_{v,\pi},t_{v,0})$, alors l'exposant de
Liapounoff est essentiellement constant sur chacun des intervalles
$\check I_\nu$. Par contre, si $|E_\pi-E_0|\ll\max(t_{v,\pi},t_{v,0})$,
alors, sur cet ensemble exponentiellement petit, la variation de
l'exposant de Liapounoff peut {\^e}tre de l'ordre d'une constante. Prenons
un exemple. Supposons que $t_{v,0}\ll t_{v,\pi}$, ou, plus exactement,
qu'il existe $\delta>0$ tel que
\begin{equation*}
  \forall E\in V_*\cap J,\quad S_{v,0}(E)>S_{v,\pi}(E)+\delta.
\end{equation*}
Si $E_0$ et $E_\pi$ co{\"\i}ncident, alors, $\check I_0\subset \check
I_\pi$, et, au centre de l'intervalle $\check I_\pi$, l'exposant de
Liapounoff vaut
\begin{equation*}
    \Theta(E,\varepsilon)=\frac{\varepsilon}{\pi}\log\tau+o(1)=
    \frac{1}{2\pi}\left(S_h(\bar E)-S_{v,\pi}(\bar E)-S_{v,0}(\bar
    E)\right)+o(1).
\end{equation*}
Alors qu'au bord de $\check I_\pi$, il vaut
\begin{equation*}
  \begin{split}
    \Theta(E,\varepsilon)&=\frac{\varepsilon}{\pi}\log\tau+
    \frac{\varepsilon}{2\pi}\log(t_{v,\pi}(\bar E)/t_{v,0}(\bar
    E))+o(1) \\&=\frac{1}{2\pi}\left(S_h(\bar E)-2S_{v,\pi}(\bar
      E)\right)+o(1).
  \end{split}
\end{equation*}
L'exposant de Liapounoff varie donc essentiellement de
$\frac{1}{2\pi}\,(S_{v,0}(\bar E)-S_{v,\pi}(\bar E))$ sur l'intervalle
de longueur exponentiellement petite. On observe donc une chute
brutale de l'exposant de Liapounoff sur l'intervalle contenant du
spectre quand on part d'un des bord de $\check I_\pi$ pour rejoindre
le point $E_\pi$.
\subsubsection{Quand $\tau$ est petit}
\label{sec:quand-tau-p}
On suppose maintenant que $\tau\ll1$, c'est-{\`a}-dire que~\eqref{eq:14}
est v{\'e}rifi{\'e}e. Dans ce cas, le comportement spectral d{\'e}pend de la
valeur de la constante $\Lambda_n(V)$ d{\'e}finie dans la
section~\ref{sec:le-coefficient-theta}. Remarquons que $\Lambda_n(V)$
ne d{\'e}pend que de $V$. Dans cet article, on suppose que
\begin{equation}
  \label{eq:41}
  \Lambda_n(V)>1.
\end{equation}
D'apr{\`e}s le Th{\'e}or{\`e}me~\ref{w:prop}, cette condition est v{\'e}rifi{\'e}e
pour un potentiel $V$ g{\'e}n{\'e}rique .\\
Il y plusieurs \og sc{\'e}narios \fg\ possibles pour le comportement spectral
dans le cas $\tau\ll1$. Avant de les d{\'e}tailler, commen{\c c}ons par une
description g{\'e}n{\'e}rale du lieu et de la nature du spectre. On d{\'e}montre
le
\begin{Th}
  \label{thr:5}
  Pla{\c c}ons nous dans les conditions du Th{\'e}or{\`e}me~\ref{thr:2}. Supposons
  que~\eqref{eq:14} et~\eqref{eq:41} sont v{\'e}rifi{\'e}es. Alors, il existe
  $\varepsilon_0>0$ tel que, pour $\varepsilon\in]0,\varepsilon_0[$,
  et $E_0$ et $E_\pi$, deux {\'e}nergies r{\'e}sonantes dans $V^*\cap J$
  telles que l'in{\'e}galit{\'e}~\eqref{eq:20} est v{\'e}rifi{\'e}e, le spectre de
  $H_{\zeta,\varepsilon}$ dans $I_0^{(l)}\cup I_\pi^{(l')}$ est
  contenu dans l'ensemble $\Sigma(\varepsilon)$ des {\'e}nergies $E$
  v{\'e}rifiant
  \begin{equation}
    \label{eq:37}
    \left|\tau^2\xi_0(E)\xi_\pi(E)+2\Lambda_n(V)\right|\leq
    \left(2+\tau^2|\xi_0(E)|+\tau^2|\xi_\pi(E)|\right)(1+ o(1)).
  \end{equation}
  o{\`u} $o(1)$ tend vers $0$ quand $\varepsilon$ tend vers $0$ et ne
  d{\'e}pend que de $\varepsilon$.
\end{Th}
\noindent On peut pr{\'e}ciser ce th{\'e}or{\`e}me gr{\^a}ce {\`a} la
\begin{Pro}
  \label{pro:1}
  L'ensemble $\Sigma(\varepsilon)$ d{\'e}fini dans le Th{\'e}or{\`e}me~\ref{thr:5}
  est la r{\'e}union de deux intervalles ferm{\'e}s disjoints.
\end{Pro}
\noindent Ce r{\'e}sultat est une cons{\'e}quence imm{\'e}diate de~\eqref{eq:37}, des
d{\'e}finitions de $\xi_0$ et $\xi_\pi$, et des faits simples suivants :
\begin{enumerate}
\item $\Lambda_n(V)>1$ ;
\item $E\mapsto\xi_0(E)$ et $E\mapsto\xi_\pi(E)$ sont lin{\'e}aires en $E$
  ;
\item le produit $\xi_0(E)\xi_\pi(E)$ est positif pour $E$ dans
  l'intervalle $]E_0,E_\pi[$.
\end{enumerate}
On note les intervalles d{\'e}crits dans la Proposition~\ref{pro:1} par
$I_0$ et $I_\pi$.\\
On montre le
\begin{Th}
  \label{thr:6}
  Pla{\c c}ons nous dans les conditions du Th{\'e}or{\`e}me~\ref{thr:5}.
  Si $dN_{\varepsilon}(E)$ est la mesure de densit{\'e} d'{\'e}tats
  $H_{\zeta,\varepsilon}$, alors
  \begin{equation*}
    \int_{I_0}dN_{\varepsilon}(E)=\int_{I_\pi}dN_{\varepsilon}(E)
    =\frac\varepsilon{2\pi}.
  \end{equation*}
\end{Th}
\noindent Ce th{\'e}or{\`e}me implique que chacun des intervalles
$I_0$ et $I_\pi$ contient du spectre. Cela et la
Proposition~\ref{pro:1} montre que, dans le cas o{\`u} $\tau\ll1$ et
$\Lambda_n(V)>1$, il y a une r{\'e}pulsion (\og splitting \fg) des intervalles
r{\'e}sonants contenant du spectre.
\smallpagebreak Pour ce qui concerne la nature du spectre, on d{\'e}montre
les r{\'e}sultats suivants. Le comportement de l'exposant de Liapounoff
est d{\'e}crit par le
\begin{Th}
  \label{thr:7}
  Pla{\c c}ons nous dans les conditions du Th{\'e}or{\`e}me~\ref{thr:5}. Dans
  l'ensemble $\Sigma(\varepsilon)$, l'exposant de Liapounoff de
  $H_{\zeta,\varepsilon}$ v{\'e}rifie
  \begin{equation}
    \label{eq:40}
    \Theta(E,\varepsilon)=\frac{\varepsilon}{\pi}
    \log\left(\tau\sqrt{|\xi_0(E)|+|\xi_\pi(E)|}\right)+o(1),
  \end{equation}
  o{\`u} $o(1)$ tend vers $0$ quand $\varepsilon$ tend vers $0$ et ne
  d{\'e}pend que de $\varepsilon$.
\end{Th}
\noindent Pour $c>0$, on d{\'e}finit l'ensemble
\begin{equation}
  \label{eq:36}
  I^+_c=\left\{E\in \Sigma(\varepsilon):\ \frac{\varepsilon}{\pi}
    \log\left(\tau\sqrt{|\xi_0(E)|+|\xi_\pi(E)|}\right)>c\right\}.
\end{equation}
Le Th{\'e}or{\`e}me~\ref{thr:7} implique le
\begin{Cor}
  \label{cor:1}
  Pla{\c c}ons nous dans les conditions du Th{\'e}or{\`e}me~\ref{thr:5}. Pour
  $\varepsilon$ suffisamment petit, l'ensemble
  $\Sigma(\varepsilon)\cap I^+_c$ ne contient que du spectre
  singulier.
\end{Cor}
\smallpagebreak Soit
\begin{equation}
  \label{eq:34}
  I^-_c=\left\{E\in\Sigma(\varepsilon):\ \frac{\varepsilon}{\pi}
    \log\left(\tau\sqrt{|\xi_0(E)|+|\xi_\pi(E)|}\right)<-c\right\}.
\end{equation}
On d{\'e}montre le
\begin{Th}
  \label{thr:1}
   Pla{\c c}ons nous dans les conditions du Th{\'e}o\-r{\`e}me~\ref{thr:5}.
   Alors, il existe $\eta>0$, une constante, et $D\subset (0,1)$,
   un ensemble de nombres diophantiens, tels que
  \begin{itemize}
  \item $D$ est asymptotiquement de mesure totale i.e.
    \begin{equation*}
      \frac{\mes(D\cap(0,\varepsilon))}{\varepsilon}=
      1+o\left(e^{-\eta/\varepsilon}\right)\text{
        lorsque }\varepsilon\to0.
    \end{equation*}
  \item pour $\varepsilon\in D$ suffisamment petit, l'ensemble
    $\Sigma(\varepsilon)\cap I^-_c$ contient majoritairement du
    spectre absolument continu de $H_{\zeta,\varepsilon}$ ; plus
    pr{\'e}cis{\'e}ment, si $\Sigma_{ac}$ d{\'e}signe le spectre absolument
    continu de $H_{\zeta,\varepsilon}$, on a
    \begin{equation*}
      \frac{\mes([\Sigma(\varepsilon)\cap I^-_c]\cap \Sigma_{\rm ac})}
        {\mes(\Sigma(\varepsilon)\cap I^-_c)}=1+o(1)
    \end{equation*}
    o{\`u} $o(1)$ tend vers $0$ quand $\varepsilon$ tend vers $0$.
  \end{itemize}
\end{Th}
\subsubsection{Les principaux sc{\'e}narios possibles quand $\tau$ est petit}
\label{sec:les-scen-princ}
Comme dans la section pr{\'e}c{\'e}\-dente, nous supposons que $\tau\ll1$ et
$\Lambda_n(V)>1$. Il y a principalement deux comportements possibles
pour le lieu et la nature du spectre. D{\'e}finissons le param{\`e}tre
\begin{equation}
  \label{eq:1}
  \rho:=\left.\frac{\max(t_{v,\pi},t_{v,0})}{\sqrt{t_h}}\right|_{E=\bar E}.
\end{equation}
Nous ne consid{\'e}rerons en d{\'e}tails que les cas quand $\rho$ est
exponentiellement petit et quand $\rho$ est exponentiellement grand.
\smallpagebreak{\bf 1.} \ Lorsque $\rho\ll1$ et $|E_\pi-E_0|\ll
\varepsilon\sqrt{t_h(\bar E)}$, $\Sigma(\varepsilon)$ consiste en deux
intervalles de longueur d'ordre $\varepsilon\sqrt{t_h}$ qui sont
s{\'e}par{\'e}s par une lacune de longueur, elle aussi, d'ordre
$\varepsilon\sqrt{t_h}$ (voir la figure~\ref{fig:rho_petit}) ; cette
lacune est essentiellement \og centr{\'e}e \fg\ au point $\bar E$.  La
longueur des intervalles contenant le spectre ainsi que la longueur et
le centre de la lacune ne d{\'e}pendent que tr{\`e}s marginalement de la
distance $E_\pi-E_0$. Lorsque $|E_\pi-E_0|$ approche
$\varepsilon\sqrt{t_h(\bar E)}$, les intervalles contenant le spectre
commencent {\`a} s'{\'e}loigner l'un de l'autre.
%
%%%%%%%%%%%%%%%%%%%%%%%%%%%%%%%%%%%%%%%%%%%%%%%%%%%%%%%%%%%%%%%%%%%%%% %
\begin{floatingfigure}[r]{6cm}
  \centering
    \includegraphics[bbllx=71,bblly=655,bburx=275,bbury=721,width=6cm]{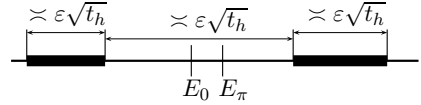}
    \caption{Quand $\tau\ll1$ et $\rho\ll1$}\label{fig:rho_petit}
\end{floatingfigure}
%%%%%%%%%%%%%%%%%%%%%%%%%%%%%%%%%%%%%%%%%%%%%%%%%%%%%%%%%%%%%%%%%%%%% %
%
\par
\begin{figure}[htbp]
  \centering
  \subfigure[Quand $|E_\pi-E_0|\gg\varepsilon t_{v,\pi}(\bar E)$]{
    \includegraphics[bbllx=56,bblly=655,bburx=275,bbury=721,width=7cm]{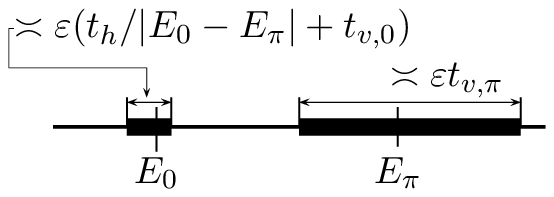}
    \label{fig:rho_grand0}}
  \hskip1cm
  \subfigure[Quand $|E_\pi-E_0|\ll\varepsilon t_{v,\pi}(\bar E)$]{
    \includegraphics[bbllx=52,bblly=655,bburx=261,bbury=721,width=7cm]{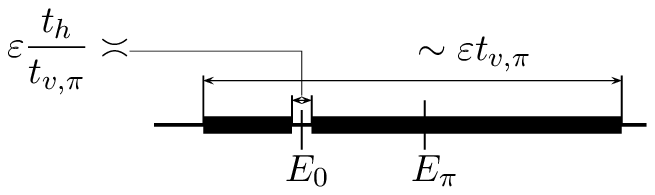}
    \label{fig:rho_grand1}}
  \caption{Le lieu du spectre quand $\tau\ll1$ et $\rho\gg1$}
  \label{fig:tau_petit}
\end{figure}

 Pour ce qui concerne la nature du spectre, lorsque $\rho$ est
exponentiellement petit et que $|E_\pi-E_0|\ll \varepsilon\sqrt{
  t_h(\bar E)}$, les intervalles contenant du spectre se trou\-vent
dans la zone $I^-_c$ ; ainsi, dans ce cas, l'essentiel du spectre est
absolument continu dans les deux intervalles (si $\varepsilon$ v{\'e}rifie
la condition diophantienne du Th{\'e}o\-r{\`e}me~\ref{thr:1}).
\vskip.2cm\noindent{\bf 2.} \ Si $\rho\gg1$, il y a toujours un
intervalle, disons $I_\pi$, qui est asymptotiquement centr{\'e} en $\pi$
et qui contient du spectre. La longueur de cet intervalle et d'ordre
$\varepsilon t_{v,\pi}(\bar E)$.\\
On peut distinguer deux cas :
\begin{enumerate}
\item si $E_0$ se trouve {\`a} l'int{\'e}rieur de $I_ \pi$ et si la distance
  entre $E_0$ et les bords de $I_\pi$ est de m{\^e}me ordre que la
  longueur de $I_\pi$, alors $\Sigma(\varepsilon)$ consiste en
  l'intervalle $I_\pi$ priv{\'e} d'une lacune de longueur d'ordre
  $\varepsilon t_h(\bar E)/t_{v,\pi}(\bar E)$ contenant $E_0$ (voir la
  figure~\ref{fig:rho_grand1}). De plus, la distance de $E_0$ {\`a} chacun
  des bords de la lacune est aussi d'ordre $\varepsilon t_h(\bar
  E)/t_{v,\pi}(\bar E)$.
\item si $E_0$ se trouve {\`a} l'ext{\'e}rieur de $I_ \pi$ et la distance
  entre $E_0$ et $I_\pi$ est au moins de m{\^e}me ordre que la longueur de
  $I_\pi$, alors $\Sigma(\varepsilon)$ consiste en l'intervalle
  $I_\pi$ et un intervalle $I_0$ (voir la
  figure~\ref{fig:rho_grand0}). L'intervalle $I_0$ est situ{\'e} dans un
  voisinage de $E_0$ dont la taille est d'ordre $\varepsilon t_h(\bar
  E)/|E_0-E_\pi|$.  La longueur de $I_0$ est d'ordre $\varepsilon(
  t_h(\bar E)/|E_0-E_\pi|+ t_{v,0}(\bar E))$.
\end{enumerate}
%
%%%%%%%%%%%%%%%%%%%%%%%%%%%%%%%%%%%%%%%%%%%%%%%%%%%%%%%%%%%%%%%%%%%%%% %
%%%%%%%%%%%%%%%%%%%%%%%%%%%%%%%%%%%%%%%%%%%%%%%%%%%%%%%%%%%%%%%%%%%%%% %
%
Lorsque $\rho$ est exponentiellement grand, le Liapounoff peut varier
tr{\`e}s brutalement sur les intervalles contenant du spectre.
Consid{\'e}rons le cas o{\`u} $|E_\pi-E_0|\ll\varepsilon t_{v,\pi}(\bar E)$.
Pour $E$ pr{\`e}s de la lacune entourant $E_0$, on voit que
$\tau^2\xi_0(E)$ est d'ordre $1$, alors que $\tau^2\xi_\pi(E)$ est
exponentiellement petit.  Donc, pour $E$ pr{\`e}s de la lacune entourant
$E_0$, le Th{\'e}or{\`e}me~\ref{thr:7} donne $\Theta(E,\varepsilon)=o(1)$. Par
contre, aux autres extr{\'e}mit{\'e}s de l'intervalle contenant le spectre,
$\tau^2\xi_0(E)$ est de taille $\rho^2$ ; comme ce dernier est
exponentiellement grand en $\varepsilon$, on voit que, pour de telles
{\'e}nergies, l'exposant de Liapounoff est positif et donn{\'e} par
\begin{equation*}
  \Theta(E,\varepsilon)=\frac{\varepsilon}{\pi}\log\rho+o(1).
\end{equation*}
Cette situation est semblable {\`a} celle obtenue pour $\tau\gg1$ ; mais,
maintenant, l'exposant varie brutalement entre une valeur positive et
une valeur tr{\`e}s petite qui pourrait {\^e}tre nulle.\\
Sur la plus grande partie de $\Sigma(\varepsilon)$, l'exposant de
Liapounoff est positif et, le spectre est singulier (d'apr{\`e}s le
Corollaire~\ref{cor:1}). Par contre, pr{\`e}s de la lacune entourant
$E_0$, ni le Corollaire~\ref{cor:1}, ni le Th{\'e}or{\`e}me~\ref{thr:1} ne
s'applique. Ces zones sont semblables aux zones o{\`u} ont lieu les
transitions d'Anderson asymptotiques (voir~\cite{MR2003f:82043}).
\subsubsection{Quand $\tau$ n'est ni petit, ni grand}
\label{sec:quand-tau-e}
Supposons qu'il existe une constante $C>1$ et une zone d'{\'e}nergies $Z$
tels que
\begin{equation}
  \label{model-cond}
   \forall E\in Z,\quad \frac1C\le\tau(E)\le C\quad\text{et}
   \quad|\xi_0(E)|+|\xi_\pi(E)|\le C.
\end{equation}
Dans ce cas, on pose $h=\varepsilon\cdot (\frac{2\pi}{\varepsilon}\,{\rm
  mod}\, 1)$, et l'{\'e}tude spectrale de $H_{\zeta,\varepsilon}$ se
ram{\`e}ne {\`a} l'{\'e}tude des solutions de l'{\'e}quation aux diff{\'e}rences finies
\begin{equation}
  \label{model-eq}
  \Psi_{k+1}=M(E,kh+\zeta) \Psi_k,\quad \Psi_k\in \C^2,\quad k\in\Z,
\end{equation}
o{\`u} la matrice $M(E,\zeta)$ satisfait
\begin{equation}
  \label{model-mat}
  M(E,\zeta)=\begin{pmatrix}
    \tau^2g_0(\zeta)g_\pi(\zeta)+\theta^{-1}_n &\tau g_0(\zeta)
    \\{}\\\theta_n \tau g_\pi(\zeta) & \theta_n
  \end{pmatrix}+o(1),
\end{equation}
et $\theta_n$ est la solution de l'{\'e}quation $2\Lambda_n=\theta_n+
\theta^{-1}_n$ dans $[1,+\infty[$ ; en outre, on a pos{\'e}
\begin{equation*}
  g_\nu=\xi_\nu+\sin(2\pi\zeta/\varepsilon+\phi_\nu),\quad
  \nu\in\{0,\pi\}.
\end{equation*}
Ici, $\phi_0$ et $\phi_\pi$ sont des fonctions r{\'e}elles analytiques de
$E$ ind{\'e}pendantes de $\zeta$. Dans~\eqref{model-mat}, $o(1)$ d{\'e}signe
une fonction {\`a} valeurs matricielles, r{\'e}elle analytique de $E$ et
$\zeta$.  Elle est $\varepsilon$-p{\'e}riodique en $\zeta$ et
exponentiellement petite quand $\varepsilon$ tend vers $0$.\\
Le comportement des solutions de l'{\'e}quation~\eqref{model-eq} imite le
comportement des solutions de l'{\'e}quation
$H_{\zeta,\varepsilon}\psi=E\psi$ au
sens du Th{\'e}or{\`e}me 2.1 de~\cite{MR2003f:82043}.\\
Sous les conditions~\eqref{model-cond}, le terme principal de la
matrice~\eqref{model-mat} ne contient plus de param{\`e}tre asymptotique.
L'{\'e}quation~\eqref{model-eq} o{\`u} la matrice est remplac{\'e}e par son terme
principal est une {\'e}quation mod{\`e}le, le terme principal de la matrice
$M$ jouant le r{\^o}le de \og hamiltonien effectif \fg. On se trouve dans un
r{\'e}gime analogue {\`a} celui de la transition d'Anderson asymptotique
obtenu dans~\cite{MR2003f:82043}.
\begin{Rem}
  Si l'une des \og variables \fg\ $\xi_\nu(E)$ devient grande ou si le
  param{\`e}tre $\tau$ devient soit grand soit petit, alors, de nouveau,
  le spectre peut {\^e}tre analys{\'e} avec la m{\^e}me pr{\'e}cision que dans les
  sections pr{\'e}c{\'e}dentes.
\end{Rem}
\section{Le diagramme de phase : des r{\'e}sultats num{\'e}riques}
\label{sec:phase-diagram}
\begin{figure}[h]
  \centering
  \includegraphics[bbllx=71,bblly=312,bburx=767,bbury=722,width=14cm]{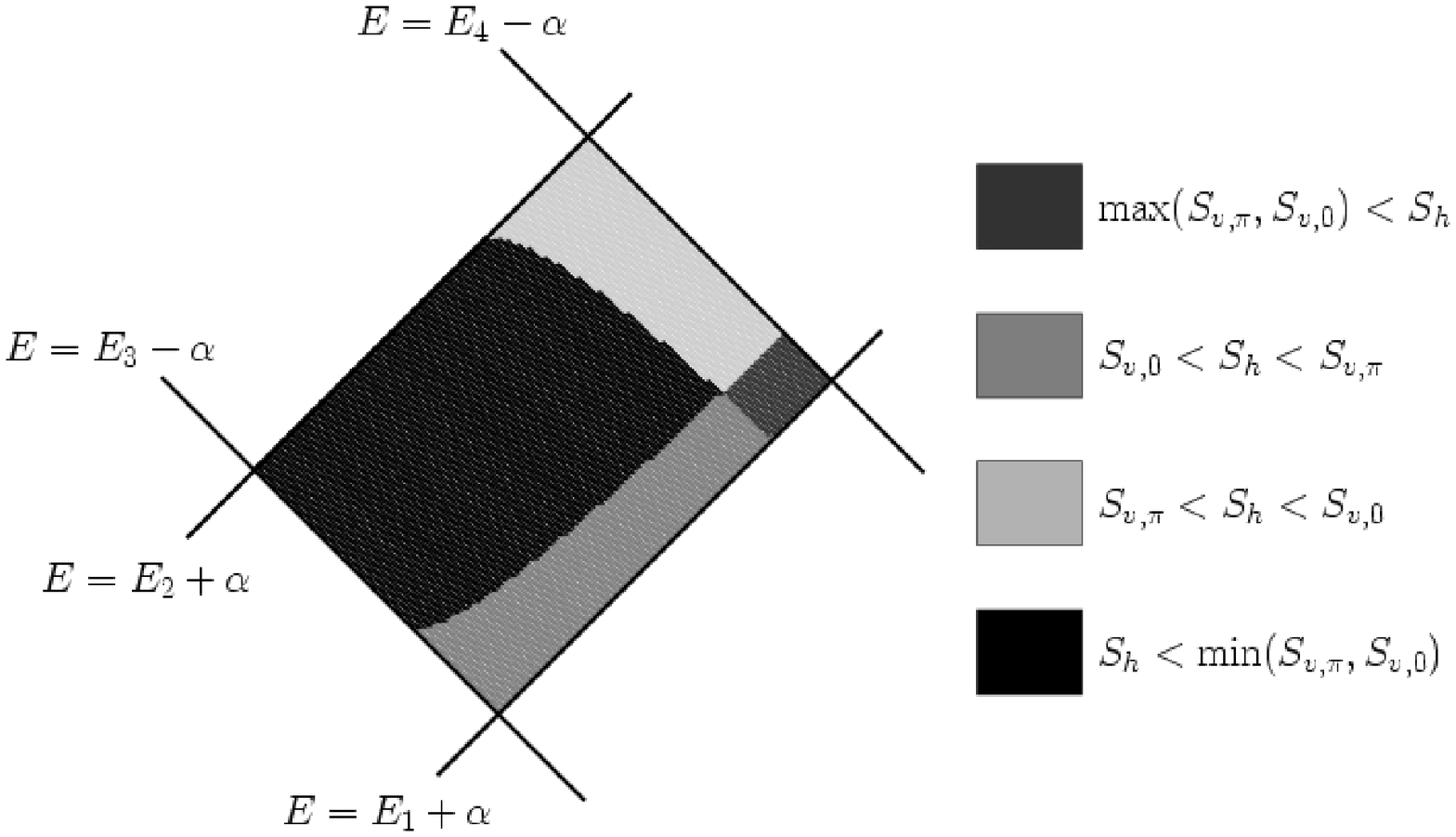}
  \caption{Comparaison des actions}
  \label{fig:action}
\end{figure}

\begin{figure}[h]
  \centering
  \includegraphics[bbllx=71,bblly=309,bburx=716,bbury=722,width=14cm]{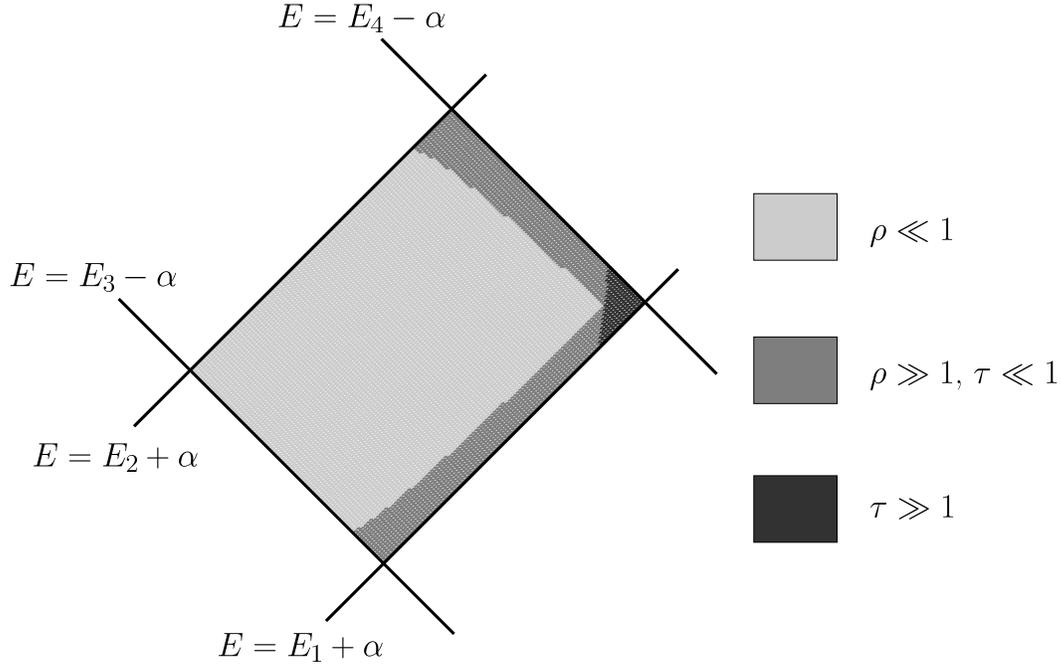}
  \caption{Comparaison des coefficients $\tau$ et $\rho$ {\`a} $1$}
  \label{fig:tau}
\end{figure}

Comme nous l'avons vu dans les th{\'e}or{\`e}mes {\'e}nonc{\'e}s pr{\'e}c{\'e}demment, les
modifications de la nature du spectre cons{\'e}cutives aux ph{\'e}nom{\`e}nes de
r{\'e}sonances d{\'e}pendent des valeurs des fonctions $S_h$, $S_{v,0}$,
$S_{v,\pi}$ sur les intervalles d'{\'e}nergies consid{\'e}r{\'e}s.\\
Bien que ces fonctions ne soient sans doute pas accessibles {\`a} un
calcul direct, pour des potentiels $V$ particuliers, on peut les
calculer num{\'e}riquement assez facilement.
\smallpagebreak Dans les r{\'e}sultats (num{\'e}riques) que nous pr{\'e}sentons
maintenant, nous avons choisi pour $V$ un potentiel {\`a} deux lacunes ;
pour de tels potentiels, on sait que la quasi-impulsion de Bloch $k$
(voir la section~\ref{sec:le-quasi-moment}) est donn{\'e}e par une
int{\'e}grale hyper-elliptique (\cite{MR52:11181,Ke-Mo:75}). Ceci rend les
actions particuli{\`e}rement faciles {\`a} calculer. Comme le spectre de
$H_0=-\Delta+V$ n'a que deux lacunes, on {\'e}crit
$\sigma(H_0)=[E_1,E_2]\cup[E_3,E_4]\cup[E_5,+\infty[$. Dans les
exemples ci-dessous, on prend les valeurs suivantes
\begin{equation*}
  E_1=0,\ E_2=3.8571429,\ E_3=6.8571429,\ E_4=12.100395,\text{
    et }E_5=100.70923.
\end{equation*}
Sur les figures~\ref{fig:action} et~\ref{fig:tau}, on a repr{\'e}sent{\'e} la
r{\'e}gion du plan des $(\alpha,E)$ o{\`u} la condition (BEI) est v{\'e}rifi{\'e}e
pour $n=1$. Elle est d{\'e}limit{\'e}e par les quatre droites d'{\'e}quation
$E=E_1+\alpha$, $E=E_2+\alpha$, $E=E_3-\alpha$ et
$E=E_4-\alpha$. Notons la $\Delta$.\\
Le calcul num{\'e}rique permet de v{\'e}rifier l'hypoth{\`e}se~(T). Comme nous
avons pris $n=1$, on a $E_{2n-2}=-\infty$. Il ne reste donc qu'{\`a}
v{\'e}rifier~(T) pour $\zeta_{2n+3}=\zeta_5$. Le fait que (T) soit vraie
peut alors {\^e}tre compris comme une cons{\'e}quence du fait que $E_5-E_4$
est grand.\\
%
%
%%%%%%%%%%%%%%%%%%%%%%%%%%%%%%%%%%%%%%%%%%%%%%%%%%%%%%%%%%%%%%%%%%%%%
%% on utilise le fichier
%% actions_nouv_zone_tibm_E2=3.8571429_E3=6.8571429_2003-04-02_041241_1.dat
%%
%%%%%%%%%%%%%%%%%%%%%%%%%%%%%%%%%%%%%%%%%%%%%%%%%%%%%%%%%%%%%%%%%%%%%
%
Sur la figure~\ref{fig:action}, on a repr{\'e}sent{\'e} les zones de la r{\'e}gion
$\Delta$ o{\`u} l'on a les divers r{\'e}gimes asymptotiques entre les
coefficients de tunnel ; dans la l{\'e}gende, ces zones sont d{\'e}crites avec
les actions. On voit que, pour les intervalles non r{\'e}sonants,
\begin{itemize}
\item les zones o{\`u} on a l'alternance de types spectraux d{\'e}crit dans le
  paragraphe~\ref{sec:alternance-des-types} ont lieu dans les zones
  $\{S_{v,0}<S_h<S_{v,\pi}\}$ et $\{S_{v,\pi}<S_h<S_{v,0}\}$
\item les transitions dues {\`a} l'approche de la situation r{\'e}sonantes
  d{\'e}crite dans le paragraphe~\ref{sec:trans-dues-lappr} ont lieu dans
  la zone $\{S_h>\max(S_{v,\pi},S_{v,0})\}$ lorsqu'on est suffisamment
  pr{\`e}s de la zone $\{S_h<\min(S_{v,\pi},S_{v,0})\}$.
\end{itemize}
Sur la figure~\ref{fig:tau}, on a repr{\'e}sent{\'e} les zones sur lesquelles
$\tau$ et $\rho$ sont grands et/ou petits.
%
%%%%%%%%%%%%%%%%%%%%%%%%%%%%%%%%%%%%%%%%%%%%%%%%%%%%%%%%%%%%%%%%%%%%%
%%%%%%%%%%%%%%%%%%%%%%%%%%%%%%%%%%%%%%%%%%%%%%%%%%%%%%%%%%%%%%%%%%%%%
%
On voit ainsi que, pour des valeurs de $\alpha$ bien choisies, tous
les ph{\'e}nom{\`e}nes d{\'e}crits dans les
paragraphes~\ref{sec:quand-tau-g},~\ref{sec:quand-tau-p}
et~\ref{sec:les-scen-princ}, c'est-{\`a}-dire dans les
figures~\ref{fig:tau_grand},~\ref{fig:rho_petit}
et~\ref{fig:tau_petit}, ont bien lieu.
\def\cprime{$'$} \def\cydot{\leavevmode\raise.4ex\hbox{.}}


\begin{thebibliography}{10}

\bibitem{Av-Si:83}
J.~Avron and B.~Simon.
\newblock Almost periodic {Schr{\"o}dinger} operators, {II}. the integrated
  density of states.
\newblock {\em Duke Mathematical Journal}, 50:369--391, 1983.

\bibitem{Bu-Fe:96}
V.~Buslaev and A.~Fedotov.
\newblock Bloch solutions of difference equations.
\newblock {\em St Petersburg Math. Journal}, 7:561--594, 1996.

\bibitem{Eas:73}
M.~Eastham.
\newblock {\em The spectral theory of periodic differential operators}.
\newblock Scottish Academic Press, Edinburgh, 1973.

\bibitem{Fe-Kl:04a}
A.~Fedotov and F.~Klopp.
\newblock On the interaction of two spectral bands of periodic {Schr{\"o}dinger}
  operator through an adiabatic incommensurate periodic perturbation: the
  non-resonant case.
\newblock In progress.

\bibitem{Fe-Kl:04b}
A.~Fedotov and F.~Klopp.
\newblock On the interaction of two spectral bands of periodic {Schr{\"o}dinger}
  operator through an adiabatic incommensurate periodic perturbation: the
  resonant case.
\newblock In progress.

\bibitem{Fe-Kl:98c}
A.~Fedotov and F.~Klopp.
\newblock A complex {WKB} analysis for adiabatic problems.
\newblock {\em Asymptotic Analysis}, 27:219--264, 2001.

\bibitem{Fe-Kl:01b}
A.~Fedotov and F.~Klopp.
\newblock On the absolutely continuous spectrum of one dimensional
  quasi-periodic {S}chr{\"o}dinger operators in the adiabatic limit.
\newblock Preprint, Universit{\'e} Paris-Nord, 2001.

\bibitem{Fe-Kl:03a}
A.~Fedotov and F.~Klopp.
\newblock On the singular spectrum of one dimensional quasi-periodic
  {Schr{\"o}dinger} operators in the adiabatic limit.
\newblock To appear in Ann. H. Poincar{\'e}, 2004.

\bibitem{Fe-Kl:03e}
A.~Fedotov and F.~Klopp.
\newblock Geometric tools of the adiabatic complex {WKB} method.
\newblock To appear in Asymp. Anal, 2004.

\bibitem{MR2003f:82043}
A.~Fedotov and F.~Klopp.
\newblock Anderson transitions for a family of almost periodic {S}chr{\"o}dinger
  equations in the adiabatic case.
\newblock {\em Comm. Math. Phys.}, 227(1):1--92, 2002.

\bibitem{MR2002f:81151}
N.~E. Firsova.
\newblock On the global quasimomentum in solid state physics.
\newblock In {\em Mathematical methods in physics (Londrina, 1999)}, pages
  98--141. World Sci. Publishing, River Edge, NJ, 2000.

\bibitem{MR81j:81010}
E.~M. Harrell.
\newblock Double wells.
\newblock {\em Comm. Math. Phys.}, 75(3):239--261, 1980.

\bibitem{He-Sj:84}
B.~Helffer and J.~Sj{\"o}strand.
\newblock Multiple wells in the semi-classical limit {I}.
\newblock {\em Communications in Partial Differential Equations}, 9:337--408,
  1984.

\bibitem{MR52:11181}
A.~R. It{\cydot}s and V.~B. Matveev.
\newblock Hill operators with a finite number of lacunae.
\newblock {\em Funkcional. Anal. i Prilo\v zen.}, 9(1):69--70, 1975.

\bibitem{MR2000f:47060}
Y.~Last and B.~Simon.
\newblock Eigenfunctions, transfer matrices, and absolutely continuous spectrum
  of one-dimensional {S}chr{\"o}dinger operators.
\newblock {\em Invent. Math.}, 135(2):329--367, 1999.

\bibitem{Ke-Mo:75}
H.~McKean and P.~van Moerbeke.
\newblock The spectrum of {Hill's} equation.
\newblock {\em Inventiones Mathematicae}, 30:217--274, 1975.

\bibitem{MR55:761}
H.~P. McKean and E.~Trubowitz.
\newblock Hill's operator and hyperelliptic function theory in the presence of
  infinitely many branch points.
\newblock {\em Comm. Pure Appl. Math.}, 29(2):143--226, 1976.

\bibitem{Pa-Fi:92}
L.~Pastur and A.~Figotin.
\newblock {\em Spectra of Random and Almost-Periodic Operators}.
\newblock Springer Verlag, Berlin, 1992.

\bibitem{MR85d:35085}
B.~Simon.
\newblock Instantons, double wells and large deviations.
\newblock {\em Bull. Amer. Math. Soc. (N.S.)}, 8(2):323--326, 1983.

\end{thebibliography}
\end{document}